\documentclass[nofootinbib,preprint,superscriptaddress,prC]{revtex4-2}

\usepackage{multirow}
\usepackage{siunitx}
\usepackage{tikz}
\usetikzlibrary{patterns}
\usetikzlibrary{decorations.pathmorphing}

\usepackage{graphicx,times}
\usepackage{latexsym}
\usepackage{mathtools}

\usepackage{amsmath,amssymb,amsbsy,amsfonts}
\usepackage{array}
\usepackage{graphics}
\usepackage{xcolor}
\usepackage{cancel}
\usepackage{xspace}
\usepackage{nicefrac}
\usepackage[normalem]{ulem}
\usepackage{dsfont}

\usepackage[unicode, pdfprintscaling=None]{hyperref}
\hypersetup{
    colorlinks,
    linkcolor={red!50!black},
    citecolor={blue!50!black},
    urlcolor={blue!80!black}
}

\usepackage[capitalise]{cleveref}

\usepackage{xcolor}
\usepackage{makecell}

\newcommand{\EFT}{EFT($\cancel{\pi}$)\xspace}
\newcommand{\vect}[1]{\vec{\boldsymbol{#1}}}
\newcommand{\vectn}[1]{#1}
\newcommand{\Gc}{\mathcal{G}}
\newcommand{\Gb}{\boldsymbol{\mathcal{G}}}

\newcommand{\Gbt}{\widetilde{\boldsymbol{\mathcal{G}}}}
\newcommand{\Gct}{\widetilde{\mathcal{G}}}
\newcommand{\Kb}{\mathbf{K}}
\newcommand{\Tb}{\mathbf{T}}

\newcommand{\Bb}{\mathbf{B}}
\newcommand{\Bbt}{{\widetilde{\mathbf{B}}}}
\newcommand{\Db}{\mathbf{D}}
\newcommand{\oneb}{\tilde{\mathbf{1}}}
\newcommand{\TS}{${}^3\!S_{1}$\xspace}
\newcommand{\SSp}{${}^1\!S_{0}$\xspace}
\newcommand{\DS}{${}^2\!S_{\frac{1}{2}}$\xspace}
\newcommand{\QS}{${}^4\!S_{\frac{3}{2}}$\xspace}
\newcommand{\CG}[6]{C_{#1,#2,#3}^{#4,#5,#6}}

\newcommand{\lambdanopi}{\Lambda_{\cancel{\pi}}}
\newcommand{\Ws}{W\!s}
\newcommand{\Was}{W\!as}
\newcommand{\NNLO}{NNLO\xspace}
\newcommand{\jjvH}{{}^3\mathrm{H}}
\newcommand{\jjvHe}{{}^3\mathrm{He}}

\newcommand{\Mcb}{\boldsymbol{\mathcal{M}}}
\newcommand{\DPo}{${}^2P_{\frac{1}{2}}$\xspace}
\newcommand{\DPt}{${}^2P_{\frac{3}{2}}$\xspace}

\usepackage{orcidlink}

\usepackage{relsize}

\usepackage{etoolbox}

\makeatletter
\newcommand*{\rom}[1]{\expandafter\@slowromancap\romannumeral #1@}

\newcolumntype{H}{>{\setbox0=\hbox\bgroup}c<{\egroup}@{}}

\begin{document}

\title{Two-Body Triton Photodisintegration and Wigner-SU(4) Symmetry}

\author{Xincheng Lin\,\orcidlink{0000-0001-9068-6787}}
\email{xincheng.lin@duke.edu}
\affiliation{Department of Physics, Box 90305, Duke University, Durham, North Carolina 27708, USA}
\affiliation{Department of Physics, North Carolina State University, Raleigh, North Carolina 27607, USA}

\author{Jared Vanasse\,\orcidlink{0000-0001-5593-6971}}
\email{jvanass3@fitchburgstate.edu}
\affiliation{Fitchburg State University, 160 Pearl St., Fitchburg, MA 01420}

\begin{abstract}

We calculate the two-body triton photodisintegration cross section as a function of photon energy to next-to-next-to leading order (NNLO) in pionless effective field theory (\EFT) and show good agreement with experiment.  In addition we calculate the polarization asymmetry $R_C=-0.441(15)$ in cold neutron-deuteron capture to NNLO in \EFT, in agreement with the experimental value of $R_C=-0.42\pm 0.03$~\cite{Konijnenberg:1988gq}.  We also assess the dependence of $R_C$ on different fits of the two-nucleon magnetic currents.  Finally, we consider the impact of Wigner-SU(4) symmetry and demonstrate that starting from the Wigner-SU(4) symmetric limit and including perturbative corrections to the breaking of Wigner-SU(4) symmetry does a good job of describing two-body triton photodisintegration.
\end{abstract}

\maketitle
\newpage

\section{Introduction}

Few-nucleon systems benefit from the fact that for any given model of nuclear interactions the system can be solved exactly and thus serve as a testing ground for models of nuclear interactions.  One observable of interest is two-body triton photodisintegration ($\jjvH\gamma\to nd$).  Near the breakup threshold this observable is dominated by the magnetic dipole moment contribution ($M1$) and beyond threshold is quickly dominated by the electric dipole moment ($E1$) contribution.  The isospin mirror process, helium-3 photodisintegration ($\jjvHe\gamma \to pd$), is also of interest, for determining deuterium abundance from Big Bang and stellar nucleosynthesis for example~\cite{Bertulani:2016eru}, but is complicated by the necessary nonperturbative inclusion of Coulomb interactions near the breakup threshold.  $\jjvH\gamma \to nd$ is also related to neutron-deuteron ($nd$) capture ($nd \to \jjvH\gamma$) via time-reversal symmetry, which therefore also depends on the $M1$ and $E1$ moments.  Cold $nd$ capture was calculated previously in Ref.~\cite{Lin:2022yaf} using pionless effective field theory (\EFT).  This work builds upon Ref.~\cite{Lin:2022yaf} by calculating a polarization asymmetry in cold $nd$ capture and the $E1$ moment, in addition to the $M1$ moment of Ref.~\cite{Lin:2022yaf}, and using both the $E1$ and $M1$ moment to calculate two-body triton photodisintegration.

$\jjvH\gamma \to nd$ has been measured previously~\cite{Faul:1981zz,KOSIEK1966199} and calculated using various phenomenological potentials~\cite{Efros:1999uq,Schadow:2000ep,Skibinski:2002ak,Yuan:2002dh,Deltuva:2003xe}.  Recent experiments of this process have been hindered by more rigorous radiation safety standards for highly radioactive targets.  However, there is recent interest in performing photodisintegration experiments of tritium to investigate the neutron-neutron and neutron-proton ($np$) scattering lengths \cite{Malone:2022ohg}.

The polarization asymmetry $R_C$, in $nd\to \jjvH\gamma$, is given by polarizing the neutron in $nd$ capture and measuring the difference divided by the sum of cross sections for different outgoing photon polarizations relative to the incoming neutron polarization.  $R_C$ has been measured and found to be $R_C=-0.42\pm 0.03$~\cite{Konijnenberg:1988gq} at a neutron lab velocity of $v=2200$ m/s.  The unpolarized cold $nd$ capture cross section ($\sigma_{nd}$) has been measured and found to be $\sigma_{nd}=0.508\pm0.015$~mb at a neutron lab velocity of $v=2200$ m/s~\cite{Jurney:1982zz}.  At this low energy cold $nd$ capture is dominated by $M1$ capture, which contains two contributions from an incoming $nd$ state either in the spin doublet (\DS) or spin quartet (\QS) $S$-wave state.  $R_C$ and $\sigma_{nd}$ can be used to tease apart these two contributions from experiment and offer a more stringent test for theoretical calculations.

Although $R_C$~\cite{Viviani:1996ui,Marcucci:2005zc} and $\jjvH\gamma \to nd$~\cite{Efros:1999uq,Schadow:2000ep,Skibinski:2002ak,Yuan:2002dh,Deltuva:2003xe} have been calculated previously using potential models, these models lack any rigorous theoretical error estimation. \EFT  and effective field theories in general have the benefit of being a systematically improvable platform to calculate observables and estimate errors in theoretical calculations~\cite{vanKolck:1999mw,Hammer:2019poc}.  At low energies [i.e., single-nucleon energies $E<m_\pi^2/(2M_N)$] pions can be integrated out leaving only contact interactions between nucleons and possible external currents in \EFT.  In principle there are an infinite number of such contact interactions, but in practice these interactions are ordered by the power counting in powers of $Q\sim \widetilde{Q}/\lambdanopi\sim 1/3$, where $\widetilde{Q}$ is the external momentum of nucleons and $\lambdanopi\sim m_\pi$ the cutoff of \EFT.  By use of the power counting only a finite number of terms are necessary at each order and errors can be estimated in theoretical calculations.  \EFT has been used to great success for two-nucleon sytems~\cite{Chen:1999tn,Chen:1999vd,Rupak:1999rk}, three-nucleon systems~\cite{Bedaque:1998mb,Bedaque:1999ve,Gabbiani:1999yv,Bedaque:2002yg,Griesshammer:2004pe,Vanasse:2013sda,Rupak:2001ci,Koenig:2011lmm,Konig:2013cia,Vanasse:2014kxa,Konig:2014ufa,Konig:2016iny}, and $A>3$ systems~\cite{Platter:2004zs,Stetcu:2006ey,Kirscher:2010dgl,Kirscher:2011uc,Bansal:2017pwn,Contessi:2017rww,Kirscher:2018dwo,Konig:2019xxk,Schafer:2022hzo,Bagnarol:2023crb}.  In the three-nucleon sector \EFT was used to include external currents~\cite{Vanasse:2015fph,DeLeon:2016wyu,Vanasse:2017kgh,Kirscher:2017fqc,Konig:2019xxk,De-Leon:2020glu} and was also used to calculate cold $nd$ capture~\cite{Sadeghi:2006fc, Arani:2014qsa,Lin:2022yaf} and $\jjvH\gamma \to nd$~\cite{Sadeghi:2009rf}.  However, the calculations in Refs.~\cite{Sadeghi:2006fc,Sadeghi:2009rf} were not strictly perturbative since they made use of the so called partial resummation technique~\cite{Gabbiani:1999yv, Bedaque:2002yg} in which all the correct contributions up to a given order are included but also an infinite subset of higher order contributions.  In addition Ref.~\cite{Sadeghi:2009rf} (Refs.~\cite{Sadeghi:2006fc,Sadeghi:2009rf}) calculated the $E1$ (
$M1$) moment to LO (NNLO) but missed diagrams.  Reference~\cite{Arani:2014qsa} corrected the calculation of Refs.~\cite{Sadeghi:2006fc,Sadeghi:2009rf} for the $M1$ moment by including the missing diagrams, but did not recalculate the $E1$ moment that also suffers from missing diagrams.  This work seeks to alleviate these issues by doing the first strictly perturbative calculation of $\jjvH\gamma \to nd$ up to and including next-to-next-to leading order (\NNLO) in \EFT by calculating the $E1$ moment strictly perturbatively using techniques from Refs.~\cite{Vanasse:2013sda,Lin:2022yaf}.

Combining spin and isospin into a single complex four dimensional object one can consider so called Wigner-SU(4) symmetry~\cite{Wigner:1936dx} in which interactions are invariant under arbitrary rotations in the complex four-dimensional spin-isospin space.  Although nuclear interactions violate Wigner-SU(4) symmetry, the breaking of Wigner-SU(4) symmetry can be included perturbatively in certain systems as demonstrated in Ref.~\cite{Vanasse:2016umz,Chen:2004rq}.  Up to next-to-leading order (NLO) in \EFT Wigner-SU(4) symmetry is satisfied when the scattering lengths and effective ranges in the spin-triplet (\TS) and spin-singlet (\SSp) channels are equal.  The breaking of Wigner-SU(4) symmetry at leading order (LO) in \EFT is characterized by the parameter
\begin{equation}
\delta^*=\frac{1}{2}\left(\frac{1}{a^{^{3}S_1}}-\frac{1}{a^{^{1}S_0}}\right)\approx 27\mathrm{\,MeV} \ \ ,
\end{equation}
where $a^{^{3}S_1}$ ($a^{^{1}S_0}$) is the \TS (\SSp) channel scattering length.  Reference~\cite{Vanasse:2016umz} used an expansion in powers of $\delta^*/\kappa_3$, where $\kappa_3$ is a three-body parameter (e.g., the triton binding momentum), to calculate observables in three-nucleon bound-state systems and found good agreement with experiment.  Building on this Ref.~\cite{Lin:2022yaf} used the same expansion to analyze cold $nd$ capture.  However, this analysis was hindered by the fact that although three-nucleon bound states are described well by perturbative corrections from Wigner-SU(4) breaking $nd$ scattering states are not.\footnote{In the Wigner-SU(4) expansion the dibaryon propagators are expanded and  $\kappa_3\sim\gamma-\sqrt{\frac{3}{4}p^2-M_NE}$, where $\gamma=[(1/a^{^{3}S_1})+(1/a^{^{1}S_0})]/2$.  For bound states $E=-8.48$~MeV, the triton binding energy, and $\kappa_3\gtrsim 70$~MeV, larger than $\delta^*$.  For scattering states, for various values of $p$ and $E$, $\kappa_3$ can be quite small in comparison to $\delta^*$.}  To do a proper analysis the Wigner-SU(4) expansion should only be used on the three-nucleon states after the emission of the photon but not before the emission of the photon in $nd$ capture.  This leads to a complication of using the three-body force in the doublet $S$-wave channel consistently before and after the photon and as a result Ref.~\cite{Lin:2022yaf} did not carry out a rigorous Wigner-SU(4) expansion for $nd$ capture.  However, for the $E1$ moment the $nd$ scattering states in $nd$ capture are $P$-wave states and therefore do not have a three-body force up to and including NNLO and a consistent Wigner-SU(4) expansion for both three-nucleon bound states and $nd$ scattering states is relatively straightforward.  This work carries out this expansion and shows it does a good job describing the $E1$ moment for two-body triton photodisintegration.

This paper is organized as follows. Section~\ref{sec:Lag} gives the Lagrangian for interactions of interest, while Sec.~\ref{sec:three} reviews essential details of the three-nucleon calculation such as the three-nucleon vertex function.  In Sec.~\ref{sec:photo} details of how to calculate both the $M1$ and $E1$ moment for two-body triton photodisintegration are given.  Section~\ref{sec:Wig} discusses the Wigner-SU(4) symmetry expansion and Sec.~\ref{sec:obs} gives the physical observables of interest.  Finally, in Sec.~\ref{sec:res} we give results and conclude in Sec.~\ref{sec:conc}.  Appendices include further details of calculations and error analysis.

\section{\label{sec:Lag}Lagrangian}
Using the auxiliary field formalism the two-nucleon Lagrangian in \EFT up to and including NNLO interactions is given by
\begin{align}
 \label{eq:twoL}   \mathcal{L}_2=&\hat{N}^{\dagger}\left(iD_0+\frac{\vect{\mathbf{D}}^2}{2M_N}\right)\hat{N}+\hat{t}_i^{\dagger}\left[\Delta_t-\sum_{n=0}^{1}c_{0t}^{(n)}\left(iD_0+\frac{\vect{\mathbf{D}}^2}{4M_N}+\frac{\gamma_t^2}{M_N}\right)\right]\hat{t}_i\\\nonumber
    &+\hat{s}_a^{\dagger}\left[\Delta_s-\sum_{n=0}^{1}c_{0s}^{(n)}\left(iD_0+\frac{\vect{\mathbf{D}}^2}{4M_N}+\frac{\gamma_s^2}{M_N}\right)\right]\hat{s}_a\\\nonumber
    &+y\left[\hat{t}_i^{\dagger}\hat{N}^TP_i\hat{N}+\hat{s}_a^{\dagger}\hat{N}^T\bar{P}_a\hat{N}+\mathrm{H.c.}\right],
\end{align}
where $\hat{N}$, $\hat{t}$, and $\hat{s}$ are the nucleon, spin-triplet dibaryon, and spin-singlet dibaryon fields respectively. $P_i=\frac{1}{\sqrt{8}}\sigma_2\sigma_i\tau_2$ ($\bar{P}_a=\frac{1}{\sqrt{8}}\sigma_2\tau_2\tau_a$) projects out the spin-triplet iso-singlet (spin-singlet iso-triplet) combination of nuclei.  The covariant derivative is defined by
\begin{equation}
    D_\mu=\partial_\mu+ie\mathbf{Q}\hat{A}_\mu,
\end{equation}
where the charge operator $\mathbf{Q}=\frac{1+\tau_3}{2}$, $1$, or $1+T_3$ for the nucleon, spin-triplet dibaryon, or spin-singlet dibaryon field respectively. $T_3$ gives the $z$-component of isospin for an isotriplet field.  In the Z parametrization~\cite{Phillips:1999hh} the bound  (virtual bound) state pole in the \TS (\SSp) channel is reproduced at LO and at NLO the residue about the pole is reproduced.  Using the Z parametrization the parameters take the values~\cite{Griesshammer:2004pe}
\begin{equation}
    y^2=\frac{4\pi}{M_N}\quad,\quad \Delta_{\{t,s\}}=\gamma_{\{t,s\}}-\mu\quad,\quad c_{0\{t,s\}}^{(n)}=(-1)^n\frac{M_N}{2\gamma_{\{t,s\}}}(Z_{\{t,s\}}-1)^{n+1},
\end{equation} 
where $\gamma_t=45.7025$~MeV, $\gamma_s=-7.890$~MeV, $Z_t=1.6908$, and $Z_s=0.9015$.  The scale $\mu$ comes from using dimensional regularization with the power divergence subtraction scheme~\cite{Kaplan:1998tg,Kaplan:1998we}.

In addition to being minimally coupled, photons can also couple magnetically to nucleons through their magnetic moments described by the Lagrangian
\begin{equation}
    \label{eq:magLO}
    \mathcal{L}_{1,0}^{mag}=\frac{e}{2M_N}\hat{N}^{\dagger}(\kappa_0+\tau_3\kappa_1)\vec{\boldsymbol{\sigma}}\cdot\vect{\mathbf{B}}\hat{N},
\end{equation}
where $\kappa_0=0.43990$ ($\kappa_1=2.35295$) is the isoscalar (isovector) nucleon magnetic moment. At NLO and NNLO there are two-nucleon magnetic photon current contact interactions described by
\begin{equation}
\label{eq:magNLO}
    \mathcal{L}_{2,n}^{mag}=\left(e\frac{L_1^{(n-1)}}{2}\hat{t}^{j\dagger}\hat{s}_3\mathbf{B}_j+\mathrm{H.c.}\right)-e\frac{L_2^{(n-1)}}{2}i\epsilon^{ijk}\hat{t}_i^{\dagger}\hat{t}_j\mathbf{B}_k,
\end{equation}
where $n=1$ ($n=2$) refers to the NLO (NNLO) contribution. A discussion of two-nucleon currents beyond NNLO can be found in Ref.~\cite{Rupak:1999rk}.  Finally, at NNLO there is also a three-nucleon magnetic moment counterterm necessary for renormalization group (RG) invariance~\cite{Lin:2022yaf} given by
\begin{equation}
\label{eq:magNNLO}
    \mathcal{L}_{3,2}^{mag}=\frac{e}{2M_N}\hat{\psi}^{\dagger}(\widetilde{\kappa}_0+\widetilde{\kappa}_1\tau_3)\vec{\boldsymbol{\sigma}}\cdot\vect{\mathbf{B}}\hat{\psi},
\end{equation}
where $\hat{\psi}$ is a trimer auxiliary field, an isodoublet representing $\jjvH$ and $\jjvHe$.

At LO in \EFT there is a three-body force~\cite{Bedaque:1998kg,Bedaque:1998km,Bedaque:1999ve}, which receives corrections at higher orders to avoid refitting, and a new energy dependent three-body force at \NNLO~\cite{Bedaque:2002yg,Ji:2012nj}.  Using the trimer auxiliary field the three-body force contribution is given by the Lagrangian
\begin{equation}
    \mathcal{L}_3=\hat{\psi}^{\dagger}\left[\Omega-h_2(\Lambda)\left(i\partial_0+\frac{\nabla^2}{6M_N}-E_B\right)\right]\hat{\psi}+\sum_{n=0}^2\omega_0^{(n)}\left[\hat{\psi}^{\dagger}\sigma_i\hat{N}\hat{t}_i-\hat{\psi}^\dagger\tau_a\hat{N}\hat{s}_a+\mathrm{H.c.}\right],
\end{equation}
where $E_B=-8.481798$~MeV is the triton binding energy. For details of how this can be related to interactions purely in terms of dibaryon and nucleon fields see Ref.~\cite{Vanasse:2015fph}.

\section{\label{sec:three}Three-nucleon}

In order to calculate properties of the triton, the triton wavefunction or equivalently the triton vertex function is needed.  The LO triton vertex function and its NLO and NNLO correction are vectors in cluster configuration (c.c.) space~\cite{Griesshammer:2004pe}  given by the integral equations
\begin{align}
    \label{eq:vertex}
    &\Gb_n(E_B,p)=\oneb\delta_{n0}+\sum_{m=1}^n\mathbf{R}_m\left(E_B,p\right)\Gb_{n-m}(E_B,p)\\\nonumber
    &\hspace{5cm}+\mathbf{K}^{\frac{1}{2}}_{0\frac{1}{2},0\frac{1}{2}}(q,p,E_B)\otimes_q\Gb_{n}(E_B,q),
\end{align}
where the subscript $n=0$, $1$, or $2$ refers to the LO vertex function, the NLO correction, and the NNLO correction to the vertex function respectively.  In c.c.~space the kernel is a matrix defined by
\begin{align}
    \label{eq:scattkernel}
    &\Kb^J_{L'S',LS}(q,p,E)=
          -\frac{2\pi}{qp}Q_L\left(\frac{q^2+p^2-M_NE-i\epsilon}{qp}\right)\\\nonumber
          &\hspace{3cm}\times\left(\!\!\begin{array}{rr}
         1 & -3\\
         -3 & 1
         \end{array}\!\right)\left[\left(\begin{array}{cc}
         1 & 0\\
         0 & 1
         \end{array}\right)\delta_{S\frac{1}{2}}+\frac{1}{4}\left(\begin{array}{cc}
         1 & 0\\
         3 & 0
         \end{array}\right)\delta_{S\frac{3}{2}}\right]\Db\left(E,q\right)\delta_{LL'}\delta_{SS'},
\end{align}
and range corrections are added perturbatively through the c.c.~space matrices
\begin{align}
    \mathbf{R}_{m}(E,p)=\left(\begin{array}{cc}
    \frac{c_{0t}^{(m-1)}}{M_N}\left(\gamma_t+\sqrt{\frac{3}{4}p^2-M_NE-i\epsilon}\,\right) & 0\\
    0 & \frac{c_{0s}^{(m-1)}}{M_N}\left(\gamma_s+\sqrt{\frac{3}{4}p^2-M_NE-i\epsilon}\,\right)
    \end{array}\right).
\end{align}
The subscripts $L$ ($L'$) and $S$ ($S'$) denote the incoming (outgoing) orbital and spin angular momentum respectively, while the superscript $J$ ($J'$) denotes the incoming (outgoing) total angular momentum.  $\oneb$ is a c.c.~space vector defined by
\begin{equation}
    \oneb=\left(\!\!\begin{array}{rr}
    1\\
    -1
    \end{array}\right).
\end{equation}
The dibaryon matrix $\Db(E,p)$ is a c.c. space matrix defined by
\begin{equation}
    \Db(E,p)=\left(\begin{array}{cc}
    D_t(E,p) & 0\\
    0 & D_s(E,p)
    \end{array}\right)=\left(\begin{array}{cc}
    \frac{1}{\gamma_t-\sqrt{\frac{3}{4}p^2-M_NE-i\epsilon}} & 0 \\
    0 & \frac{1}{\gamma_s-\sqrt{\frac{3}{4}p^2-M_NE-i\epsilon}}
    \end{array}\right).
\end{equation}
$Q_L(a)$ is a Legendre function of the second kind given in terms of the Legendre polynomials $P_L(x)$ by\footnote{Note, this differs from the conventional definition of Legendre functions of the second kind by a phase of $(-1)^L$.}
\begin{equation}
    Q_L(a)=\int_{-1}^{1}dx\frac{P_L(x)}{x+a},
\end{equation}
while $\otimes_q$ is a shorthand for integration defined by
\begin{equation}
    A(q)\otimes_qB(q)=\frac{1}{2\pi^2}\int_0^{\Lambda}\!\!dqq^2A(q)B(q).
\end{equation}

The triton vertex function wavefunction renormalization is given by~\cite{Vanasse:2015fph}
\begin{align}
    \label{eq:tritonrenorm}
    &\sqrt{Z_\psi}=\sqrt{\frac{\pi}{\Sigma_0'(E_B)}}\left[\underbrace{1\vphantom{\frac{1}{2}\frac{\Sigma_1'(E_B)}{\Sigma_0'(E_B)}}}_{\mathrm{LO}}-\underbrace{\frac{1}{2}\frac{\Sigma_1'(E_B)}{\Sigma_0'(E_B)}}_{\mathrm{NLO}}\right.\\\nonumber
    &\hspace{4cm}\left.+\underbrace{\frac{3}{8}\left(\frac{\Sigma_1'(E_B)}{\Sigma_0'(E_B)}\right)^2-\frac{1}{2}\frac{\Sigma_2'(E_B)}{\Sigma_0'(E_B)}-\frac{2}{3}M_NH_2(\Lambda)\frac{\Sigma_0(E_B)^2}{\Sigma_0'(E_B)^2}}_{\mathrm{NNLO}}+\cdots\right],
\end{align}
where the functions $\Sigma_{n}(E)$ are defined by
\begin{equation}
    \label{eq:Sigmadef}
    \Sigma_n(E)=-\pi\oneb^T\Db\left(E,\vectn{q}\right)\otimes_q\Gb_{n}(E,q).
\end{equation}
The three-body force used in this work is defined by
\begin{equation}
    H^{(0)}=H_{\mathrm{LO}}(\Lambda)\quad,\quad H^{(1)}=H_{\mathrm{NLO}}(\Lambda)\quad,\quad H^{(2)}=H_{\mathrm{NNLO}}(\Lambda)+\frac{4}{3}M_Nk_0H_2(\Lambda)
\end{equation}
where $H_{\mathrm{LO}}(\Lambda)$ is the LO three-body force and $H_{\mathrm{NLO}}(\Lambda)$ and $H_{\mathrm{NNLO}}(\Lambda)$ are the NLO and NNLO corrections respectively.  $H_2(\Lambda)$ is the energy dependent three-body force that occurs at NNLO. For further details and how the three-body forces are determined see Ref.~\cite{Vanasse:2015fph}.

\section{\label{sec:photo}Two-Body Photodisintegration Amplitude}

The $E1$ and $M1$ moment for two-body triton photodisintegration include contributions from final state interactions that can be included by calculating the half off shell $nd$ scattering amplitude.  However, taking the approach of Ref.~\cite{Lin:2022yaf}, final state interactions are included through the use of an integral equation.  The LO $E1$ and $M1$ moments and their perturbative corrections are given by the integral equation
\begin{align}
    \label{eq:Tamp}
    \Tb^{[n]}_{[X]}{}_{L'S'}^{J'}(p,k)=&\frac{e}{2M_N}\Bb^{[n]}_{[X]}{}_{L'S'}^{J'}(p,k)\\\nonumber
    &+\sum_{m=1}^{n}\mathbf{R}_m\left(E,\vectn{p}\right)\Tb^{[n-m]}_{[X]}{}_{L'S'}^{J'}(p,k)\\\nonumber
    &-\pi\delta_{X1}\delta_{L'0}\delta_{S'\frac{1}{2}}\sum_{m=0}^{n} H^{(m)}\left(\begin{array}{rr}
    1 & -1\\
    -1 & 1
    \end{array}\right)\Db\left(E,q\right)\otimes_q\Tb^{[n-m]}_{[X]}{}^{J'}_{L'S'}(q,k)\\\nonumber
    &+\sum_{L'',S''}\Kb^{J'}_{L'S',L''S''}(q,p,E)\otimes_q \Tb^{[n]}_{[X]}{}_{L''S''}^{J'}(q,k),
\end{align}
where the superscript $n=0,1,$ and $2$  of $\Tb^{[n]}$ gives the LO, NLO correction, and \NNLO correction to the $E1$ and $M1$ moment respectively, while the subscript $X=0$ ($X=1$) denotes the $E1$ ($M1$) moment.  The subscripts $L'$ and $S'$ and superscript $J'$ denote the orbital, spin, and total angular momentum of the outgoing nuclear state, respectively.  Electrically coupled photons inject orbital angular momentum but no spin angular momentum.  Therefore, $E1$ photons change the incoming \DS state of the triton to an outgoing \DPo or \DPt scattering state in which three-body forces do not occur up to and including \NNLO. Magnetically coupled photons inject spin angular momentum but no orbital angular momentum and therefore change the incoming \DS state to either an outgoing \QS or \DS state, the latter which includes three-body force contributions. The momentum $p$ is the center-of-mass momentum of the outgoing $nd$ state and $k$ is the momentum of the incoming photon in the triton rest frame, the zero-recoil limit is used throughout this paper.

\begin{figure}[hbt]
    \centering
    \includegraphics[width=100mm]{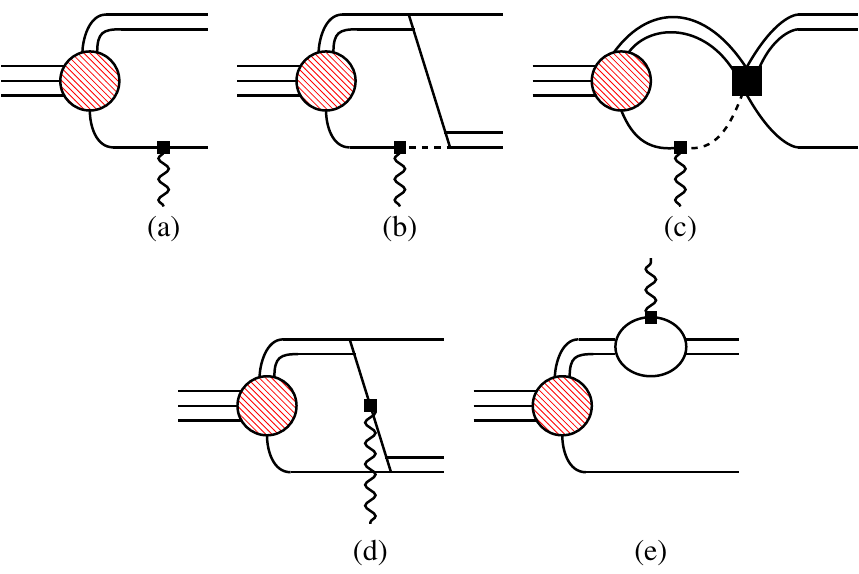}
    \caption{Diagrams for the LO inhomogeneous term, $\Bb^{[0]}_{[X]}{}_{L'S'}^{J'}(p,k)$,  of the integral equation, Eq.~\eqref{eq:Tamp}.  Shaded circles are the LO triton vertex function, single lines are nucleon propagators, wavy lines photons, and double lines are either spin-triplet or spin-singlet dibaryons.  The dashed line is a nucleon propagator whose pole is not included when integrating over energy in the loop integral.  Small black boxes attached to photons represent minimally (magnetically) coupled photons for the $E1$ ($M1$) moment and the large black box is the LO three-body force. 
    \label{fig:inhomLO}}
\end{figure}

The inhomogeneous term $\Bb^{[0]}_{[X]}{}_{L'S'}^{J'}(p,k)$ of the integral equation at LO is given by the sum of diagrams in Fig.~\ref{fig:inhomLO}
where photons are either minimally coupled, from gauging the nucleon kinetic term, for the $E1$ moment or magnetically coupled via Eq.~\eqref{eq:magLO} for the $M1$ moment.  Diagram (c) with the three-body force only occurs for the $M1$ moment. Expressions for $\Bb^{[0]}_{[X]}{}_{L'S'}^{J'}(p,k)$ are given in App.~\ref{app:simpler} for $E1$ and in Ref.~\cite{Lin:2022yaf} for $M1$.  To see diagrams in Fig.~\ref{fig:inhomLO} are of the same order, we count nucleon legs as $M_N/\widetilde{Q}^2$, the dibaryon leg as $1/\widetilde{Q}$, and the integral measure as $\dfrac{\widetilde{Q}^3\widetilde{Q}_0}{2\pi^2} \sim \dfrac{\widetilde{Q}^3\widetilde{Q}^2}{2\pi^2M_N}$, where $\widetilde{Q}$ [$\widetilde{Q}_0$] is the typical momentum [energy] scale of the system (see, e.g., Refs.~\cite{Bedaque:1998mb,Bedaque:1999ve, Gabbiani:1999yv} for details on power counting for three-nucleon bound and scattering states). Compared to diagram (a), the loops in diagrams (d) and (e) come with two additional nucleon legs, a dibaryon leg, a factor of $y^2$ from the dibaryon-nucleon interaction, and a factor of $\dfrac{\widetilde{Q}^3\widetilde{Q}_0}{2\pi^2}$ from the integral measure. This leads to a factor of 
\begin{equation}
\left(\dfrac{\widetilde{Q}^2}{M_N}\right)^{-2}\dfrac{1}{\widetilde{Q}}\dfrac{4\pi}{M_N} \dfrac{\widetilde{Q}^3 \widetilde{Q}_0}{2\pi^2} \sim \dfrac{2}{\pi} \sim 1.
\end{equation}
Therefore, diagrams (a), (d), and (e) of Fig.~\ref{fig:inhomLO} are of the same order. 
The sum of diagrams (b) and (c) can be viewed as diagram (a) multiplied by the kernel of the integral equation for \DS $nd$ scattering. Since iterations from the kernel do not change the scaling, the sum of diagrams (b) and (c) are of the same order as diagram (a).

\begin{figure}[hbt]
    \centering
    \includegraphics[width=100mm]{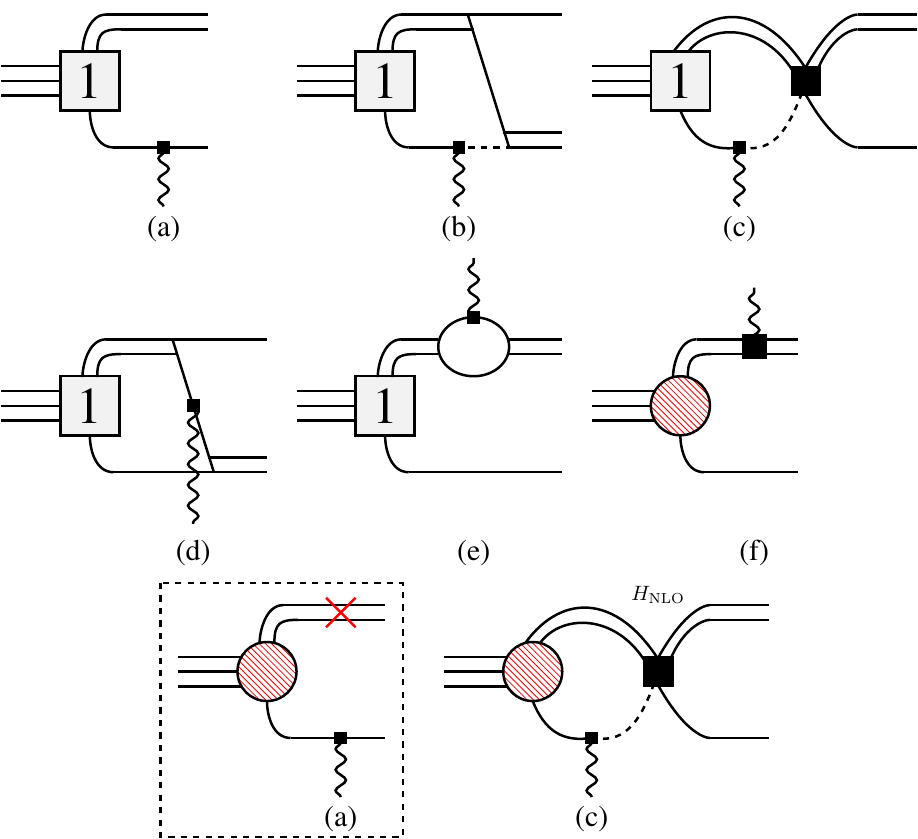}
    \caption{Diagrams for the NLO inhomogeneous term $\Bb^{[1]}_{[X]}{}_{L'S'}^{J'}(p,k)$ in Eq.~\eqref{eq:Tamp}.  The box with a ``1" is the NLO correction to the triton vertex function, the red \textcolor{red}{$\times$} represents an effective range insertion, and the black box with $H_{\mathrm{NLO}}$ is the NLO correction to the LO three-body force.  The boxed diagram is subtracted to avoid double counting between the first diagram in this figure and the $\mathbf{R}_1\Tb^{[0]}_X$ contribution to the inhomogeneous term of Eq.~\eqref{eq:Tamp}.  All other notation is the same as Fig~\ref{fig:inhomLO}.}
    \label{fig:inhomNLO}
\end{figure}

\begin{figure}[hbt]
    \centering
    \includegraphics[width=100mm]{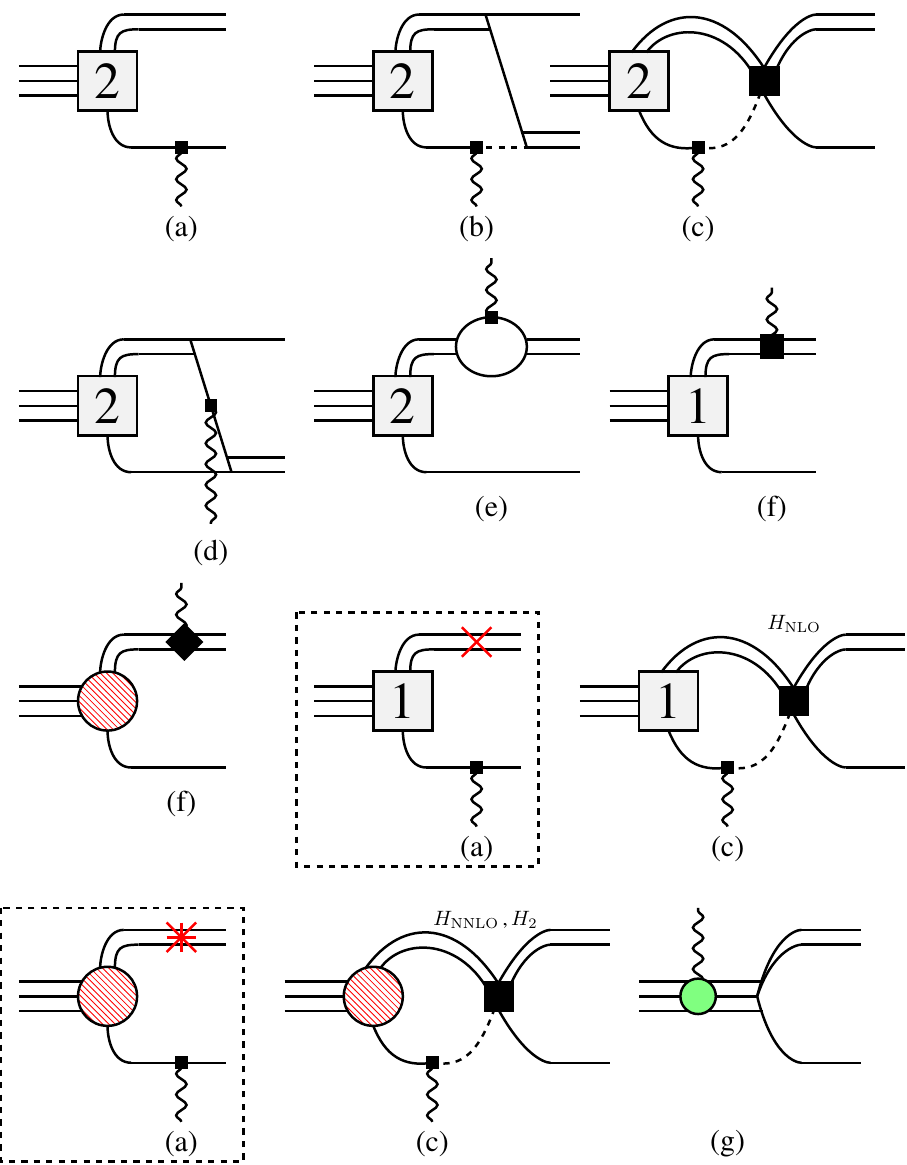}
    \caption{Diagrams for the NNLO inhomogeneous term $\Bb^{[2]}_{[X]}{}_{L'S'}^{J'}(p,k)$ in Eq.~\eqref{eq:Tamp}.  The box with a ``2" is the NNLO correction to the triton vertex function, the black diamond with a photon comes from a NNLO correction to the dibaryon kinetic term and $L_1^{(1)}$ and $L_2^{(1)}$ in Eq.~\eqref{eq:magNLO} for the $E1$ and $M1$ moments respectively.  Large black boxes represent three-body force contributions and are labeled with their respective three-body force contribution.  Boxed diagrams are subtracted to avoid double counting from the first diagram in this figure and the $\mathbf{R}_1\Tb^{[1]}_X$ and $\mathbf{R}_2\Tb^{[0]}_X$ contributions to the inhomogeneous term of Eq.~\eqref{eq:Tamp}.  All other notation is the same as Figs.~\ref{fig:inhomLO} and \ref{fig:inhomNLO}.}
    \label{fig:inhomNNLO}
\end{figure}

The NLO inhomogeneous term $\Bb^{[1]}_{[X]}{}_{L'S'}^{J'}(p,k)$ is given by the sum of diagrams in Fig.~\ref{fig:inhomNLO}
where the boxed diagram is subtracted to avoid double counting from the first diagram in Fig.~\ref{fig:inhomNLO} and the $\mathbf{R}_1\Tb^{[0]}_X$ contribution to the inhomogeneous term of Eq.~\eqref{eq:Tamp}.\footnote{Note, multiple diagrams (a) and (c) occur in Figs.~\ref{fig:inhomNLO} and~\ref{fig:inhomNNLO} as these diagrams can be naturally combined, as in Ref.~\cite{Lin:2022yaf} for the $M1$ moment.}  In diagram~(f) the photon is minimally coupled to the dibaryon through gauging its kinetic term for the $E1$ moment and it is coupled via Eq.~\eqref{eq:magNLO} for the $M1$ moment.  Finally, the NNLO inhomogeneous term $\Bb^{[2]}_{[X]}{}_{L'S'}^{J'}(p,k)$ is given by the sum of diagrams in Fig.~\ref{fig:inhomNNLO}. 
To see that the N$^m$LO diagrams, $m=1,2$, are of the same order, one can use the fact that each insertion of the effective range term gives a factor of ${Q}\sim1/3$. The N$^m$LO correction to the vertex function contains $m$ insertions of the effective range. The energy-dependent three-body force counts as $Q^2$, compared to the LO energy-independent one, and must be included for RG invariance at NNLO~\cite{Bedaque:2002yg,Ji:2012nj}. The scaling of the two-nucleon $M1$ LECs can be found using the matching in Ref.~\cite{Lin:2022yaf} and the scaling in Refs.~\cite{Chen:1999tn,Chen:1999vd,Rupak:1999rk}.  Diagram (g) in Fig.~\ref{fig:inhomNNLO} only exists for the $M1$ moment and comes from the NNLO three-nucleon magnetic moment contribution, Eq.~\eqref{eq:magNNLO}.  The three-nucleon magnetic moment counterterm is fit to reproduce the triton magnetic moment, for details of how this is done see Ref.~\cite{Lin:2022yaf}. 

Expressions for $\Bb^{[1]}_{[X]}{}_{L'S'}^{J'}(p,k)$ and $\Bb^{[2]}_{[X]}{}_{L'S'}^{J'}(p,k)$ are given in App.~\ref{app:simpler} for $E1$ and in Ref.~\cite{Lin:2022yaf} for $M1$.  To simplify the expressions for $\Bb^{[1]}_{[X]}{}_{L'S'}^{J'}(p,k)$ and $\Bb^{[2]}_{[X]}{}_{L'S'}^{J'}(p,k)$ the vertex function integral equation can be used and the definition of the integral equation can be shifted.  Details of this procedure can be found in App.~\ref{app:simpler} for the $E1$ moment and in Ref.~\cite{Lin:2022yaf} for the $M1$ moment.  Carrying out this procedure we replace the expressions in Eq.~\eqref{eq:Tamp} with
\begin{equation}
    \Bb^{[n]}_{[X]}{}_{L'S'}^{J'}(p,k)\to\widetilde{\Bb}^{[n]}_{[X]}{}_{L'S'}^{J'}(p,k)\quad,\quad \Tb^{[n]}_{[X]}{}_{L'S'}^{J'}(p,k)\to\widetilde{\Tb}^{[n]}_{[X]}{}_{L'S'}^{J'}(p,k),
\end{equation}
where $\Tb^{[n]}_{[X]}{}_{L'S'}^{J'}(p,k)$ and $\widetilde{\Tb}^{[n]}_{[X]}{}_{L'S'}^{J'}(p,k)$ are equivalent for on-shell two-body triton photodisintegration and the exact relation between them is given in App.~\ref{app:simpler} for the $E1$ moment and in Ref~\cite{Lin:2022yaf} for the $M1$ moment.  For the $E1$ moment $\widetilde{\Bb}^{[n]}_{[0]}{}_{L'S'}^{J'}(p,k)$ for an outgoing $^2P_{\frac{1}{2}}$ channel is given by
\begin{align}
    \label{eq:E1inhom}
     &\widetilde{\Bb}^{[n]}_{[0]}{}_{1\frac{1}{2}}^{\frac{1}{2}}(p,k)=-i\frac{e}{2M_N}\frac{2\tau_3}{\sqrt{3}\pi k_0}\int dqq^2\frac{1}{qp}\left[\Mcb pQ_0\left(\frac{q^2+p^2-M_NE-i\epsilon}{qp}\right)\right.\\\nonumber
    &\left.-\Mcb^T qQ_1\left(\frac{q^2+p^2-M_NE_B-i\epsilon}{qp}\right)\right]\mathbf{D}\left(E_B,q\right)\Gb_n(E_B,q)\\\nonumber
    &-i\frac{e}{2M_N}\frac{\tau_3}{2\sqrt{3} k_0}\Db^{-1}(E,p)\Mcb\Db(E_B,p)p\Gb_n(E_B,p)\\\nonumber
    &+i\frac{e}{2M_N}\frac{\tau_3}{2\sqrt{3} k_0}\sum_{m=1}^{n}\mathbf{R}_{m}\left(E,p\right)\Db^{-1}(E,p)\Mcb\Db(E_B,p)p\Gb_{n-m}(E_B,p)\\\nonumber
    &+i\frac{e}{2M_N}\frac{\tau_3}{2\sqrt{3} k_0}p\left(\Mcb\delta_{n0}\oneb+\Mcb^T\Gb_n(E_B,p)\right)\\\nonumber
    &-i\frac{e}{2M_N}\frac{\tau_3}{2\sqrt{3} k_0}\sum_{m=1}^{n}\Mcb^Tp\mathbf{R}_m\left(E_B,p\right)\Gb_{n-m}(E_B,p)\\\nonumber
\end{align}
where $k_0=k$ is the photon energy, and the c.c~space matrix $\Mcb$ is
\begin{equation}
    \label{eq:Mcb}
    \Mcb=\left(\begin{array}{rr}
    1 & -1 \\
    1 & -1
    \end{array}\right).
\end{equation}
For the triton $\tau_3$ is replaced with $-$1.  $\Bbt^{[n]}_{[0]}{}_{L'S'}^{J'}(p,k)$ for an outgoing $^2P_{\frac{3}{2}}$ channel  is given by
\begin{equation}
    \Bbt^{[n]}_{[0]}{}_{1\frac{1}{2}}^{\frac{3}{2}}(p,k)=\sqrt{2}\,\Bbt^{[n]}_{[0]}{}_{1\frac{1}{2}}^{\frac{1}{2}}(p,k).
\end{equation}
$\widetilde{\Bb}^{[n]}_{[1]}{}_{L'S'}^{J'}(p,k)$ for the $M1$ moment and outgoing \DS channel is given by~\cite{Lin:2022yaf}
\begin{align}
    \label{eq:LOinhommod}
    &\Bbt^{[n]}_{[1]}{}_{0\frac{1}{2}}^{\frac{1}{2}}(p,k)=\frac{2\tau_3\kappa_1}{\sqrt{3}k_0}\left(\!\!\begin{array}{rr}
    0 & 1\\
    -1 & 0
    \end{array}\!\!\right)\Db\left(E_B,p\right)\\\nonumber
    &\left\{(\gamma_t-\gamma_s)\Gb_n(E_B,p)\vphantom{\sum_{m=0}^{n-1}}\right.\\\nonumber
    &\hspace{1cm}\left.-\frac{1}{M_N}\sum_{m=0}^{n-1}\left[\left(c_{0t}^{(m)}-c_{0s}^{(m)}\right)(M_NE-\frac{3}{4}p^2)+c_{0t}^{(m)}\gamma_t^2-c_{0s}^{(m)}\gamma_s^2\right]\Gb_{n-1-m}(E_B,p)\right\}\\\nonumber
 &+\frac{1}{\sqrt{3}}\sum_{m=0}^{n-1}\left(\!\!\begin{array}{cc}
    -4\kappa_0c_{0t}^{(m)}-2M_NL_2^{(m)} & -2\tau_3\kappa_1c_{0s}^{(m)}-\tau_3M_NL_1^{(m)}\\
    -2\tau_3\kappa_1c_{0t}^{(m)}-\tau_3M_NL_1^{(m)} & 0
    \end{array}\!\!\right)\Db\left(E_B,p\right)\Gb_{n-1-m}(E_B,p)\\\nonumber
    &-\delta_{n2}\sqrt{3}\frac{4}{3}M_NH_2\Sigma_0(E_B)(\kappa_0-\tau_3\kappa_1)\oneb-\delta_{n2}\sqrt{3}\frac{1}{\Omega}(\widetilde{\kappa}_0(\Lambda)+\tau_3\widetilde{\kappa}_1(\Lambda))\oneb,
\end{align}
and for an outgoing \QS channel by
\begin{align}
&\Bbt^{[n]}_{[1]}{}_{0\frac{3}{2}}^{\frac{3}{2}}(p,k)=\frac{2\tau_3\kappa_1}{\sqrt{3}k_0}\left(\begin{array}{cc}
    0 & 1 \\
    0 & 0
    \end{array}\right)\Db\left(E_B,p\right)\\\nonumber
    &\left\{(\gamma_t-\gamma_s)\Gb_n(E_B,p)\vphantom{\sum_{m=0}^{n-1}}\right.\\\nonumber
    &\hspace{1cm}\left.-\frac{1}{M_N}\sum_{m=0}^{n-1}\left[\left(c_{0t}^{(m)}-c_{0s}^{(m)}\right)(M_NE-\frac{3}{4}p^2)+c_{0t}^{(m)}\gamma_t^2-c_{0s}^{(m)}\gamma_s^2\right]\Gb_{n-1-m}(E_B,p)\right\}\\\nonumber
&-\frac{1}{\sqrt{3}}\sum_{m=0}^{n-1}\left(\begin{array}{cc}
    -M_NL_2^{(m)}-2\kappa_0c_{0t}^{(m)} & \tau_3 M_NL_1^{(m)}+2\tau_3\kappa_1c_{0s}^{(m)} \\
    0 & 0
    \end{array}\right)\Db\left(E_B,p\right)\Gb_{n-1-m}(E_B,p).
\end{align}

To get the fully renormalized amplitude, $\widetilde{\Tb}^{[n]}_{[X]}{}^{J'}_{L'S'}(p,k)$ must be multiplied by appropriate factors of the triton and deuteron wavefunction renormalization yielding \footnote{For further details on the minus sign in the normalization factor consult Refs.~\cite{Lin:2022yaf,Vanasse:2015fph}} 
\begin{align}
    {M^{[n]}_{[X]}}{}^{J'}_{L'S'}(p,k)=&\left\{\sqrt{Z_\psi}_{[0]}\sqrt{Z_d}_{[0]}{\widetilde{\Tb}^{[n]}_{[X]}}{}^{J'}_{L'S'}(p,k)\right.\\\nonumber
    &+\left(\sqrt{Z_\psi}_{[1]}\sqrt{Z_d}_{[0]}-\sqrt{Z_\psi}_{[0]}\sqrt{Z_d}_{[1]}\right){\widetilde{\Tb}^{[n-1]}_{[X]}}{}^{J'}_{L'S'}(p,k)\\\nonumber
    &\left.+\left(\sqrt{Z_\psi}_{[2]}\sqrt{Z_d}_{[0]}+\sqrt{Z_\psi}_{[1]}\sqrt{Z_d}_{[1]}+\sqrt{Z_\psi}_{[0]}\sqrt{Z_d}_{[2]}\right){\widetilde{\Tb}^{[n-2]}_{[X]}}{}^{J'}_{L'S'}(p,k)\right\}^T\left(\begin{array}{c}    
    1 \\
    0
    \end{array}\right).
\end{align}
The superscript $T$ denotes the transpose of the c.c~space vector.  $\sqrt{Z_\psi}_{[n]}$ is $n$'th order correction to the triton wavefunction renormalization given in Eq.~\eqref{eq:tritonrenorm}, while 
$\sqrt{Z_d}_{[n]}$ is the $n$'th order correction to the deuteron wavefunction renormalization given by the residue about the deuteron pole for the deuteron propagator which gives
\begin{equation}
    \sqrt{Z_d}=\sqrt{\frac{2\gamma_t}{M_{n}}}\left[\underbrace{1\vphantom{\frac{1}{2}(Z_t-1)}}_{\mathrm{LO}}+\underbrace{\frac{1}{2}(Z_t-1)}_{\mathrm{NLO}}-\underbrace{\frac{1}{8}(Z_t-1)^2}_{\mathrm{NNLO}}+\cdots\right].
\end{equation}

\section{\label{sec:Wig}Quasi Wigner-SU(4) Symmetry Expansion}

In this section we study the impact of Wigner-SU(4) symmetry on $\gamma \jjvH\to nd$.  
Wigner-SU(4) symmetry is found to provide a good approximation for three-nucleon bound states~\cite{Vanasse:2016umz} but is not expected to work well for low-energy $nd$ scattering. Therefore, for $\gamma \jjvH\to nd$ we only perform a perturbative expansion in Wigner-SU(4) symmetry-breaking contributions on states before the $E1$ photon.  Our expansion here provides a consistency check on the previous study by Ref.~\cite{Vanasse:2016umz} on the Wigner-SU(4) expansion of three-nucleon bound states. As we will show in this section and Sec.~\ref{subsec:QSU(4)}, the $E1$ moment of $\gamma \jjvH\to nd$ is dominated by the Wigner-SU(4) limit contribution of the triton, consistent with the finding of Ref.~\cite{Vanasse:2016umz} that the triton is close to the Wigner-SU(4) limit. In addition, our expansion for $E1$ photons here is the first step to developing a consistent Wigner-SU(4) expansion for the $M1$ contributions in cold $nd$ capture. Such an expansion is important for understanding the power counting of cold $nd$ capture as this process is zero in the Wigner-SU(4) limit at LO in \EFT~\cite{Lin:2022yaf}.  

Perturbatively expanding in Wigner-SU(4) symmetry breaking for contributions before the $E1$ photon is equivalent to expanding around the Wigner-SU(4) limit for the triton vertex function and dibaryon propagators evaluated at $E_B$ in the inhomogeneous term of Eq.~\eqref{eq:E1inhom} and not for dibaryon propagators evaluated at scattering energies $E$. This expansion is shown explicitly in this section. The Wigner-SU(4) symmetry breaking expansion of the $\gamma\jjvH\to nd$ amplitude is then obtained perturbatively by solving the integral equation, Eq.~\eqref{eq:Tamp}, with the expanded inhomogeneous term; the kernel of the integral equation is not expanded because it comes after the photon and is evaluated at the scattering energy, $E$. Given that only contributions before the photon and not after the photon are expanded around the Wigner-SU(4) limit we term this expansion the quasi Wigner-SU(4) expansion.

\subsection{Vertex Function}
To understand the consequences of Wigner-SU(4) symmetry it is convenient to rewrite equations in the so called Wigner basis.  The vertex function in the Wigner basis is defined by~\cite{Griesshammer:2004pe}
\begin{equation}
    \Gb_{W,n}(E_B,p)=\left(\begin{array}{rr}
    1 & -1\\
    1 & 1
    \end{array}\right)\Gb_{n}(E_B,p),
\end{equation}
where
\begin{equation}
    \Gb_{W,n}(E_B,p)=\left(\begin{array}{c}
    \Gc_{\Ws,n}(E_B,p)\\
    \Gc_{\Was,n}(E_B,p)
    \end{array}\right).
\end{equation}
$\Gc_{\Ws,n}(E_B,p)$ ($\Gc_{\Was,n}(E_B,p)$) is the Wigner symmetric (antisymmetric) vertex function.  Similarly the dibaryon matrix in the Wigner basis is given by
\begin{equation}
    \Db_{W}(E,p)=\frac{1}{2}\left(\begin{array}{rr}
    1 & -1\\
    1 & 1
    \end{array}\right)\Db(E,p)\left(\begin{array}{rr}
    1 & 1\\
    -1 & 1 \\
    \end{array}\right),
\end{equation}
yielding
\begin{equation}
    \label{eq:dibmatrix}
    \Db_{W}(E,p)=\left(\begin{array}{cc}
    D_{\Ws}(E,p) & D_{\Was}(E,p)\\
    D_{\Was}(E,p) & D_{\Ws}(E,p)
    \end{array}
    \right),
\end{equation}
where $D_{\Ws}(E,p)$ ($D_{\Was}(E,p)$) is the Wigner-symmetric (Wigner-antisymmetric) dibaryon propagator defined by
\begin{equation}
    D_{\Ws}(E,p)=\frac{1}{2}(D_{t}(E,p)+D_{s}(E,p))\quad,\quad D_{\Was}(E,p)=\frac{1}{2}(D_{t}(E,p)-D_{s}(E,p)).
\end{equation}
Defining
\begin{align}
    \gamma=\frac{1}{2}(\gamma_t+\gamma_s)\quad &, \quad \rho=\left(\frac{Z_t-1}{2\gamma_t}+\frac{Z_s-1}{2\gamma_s}\right)\\
    \delta=\frac{1}{2}(\gamma_t-\gamma_s)\quad &, \quad\delta_r=\left(\frac{Z_t-1}{2\gamma_t}-\frac{Z_s-1}{2\gamma_s}\right),
\end{align}
we note as shown in Ref.~\cite{Vanasse:2016umz} that $\delta/\kappa_3^*$, where $\kappa_3^*$ is a scale associated with three-body binding, can be used as an expansion parameter. In addition
\begin{equation}
    \frac{\delta_r}{\rho}=0.095\sim Q^2.
\end{equation}
Therefore $\delta_r$ can be treated as a next-to-next-to-next-to leading order (N$^3$LO) correction and throughout this work we take the limit $\delta_r=0$. 

Carrying out an expansion in $\delta$ for the dibaryon propagators yields
\begin{equation}
    D_{\Ws}(E,p)=\sum_{n=0}^{\infty}\delta^{2n}[D(E,p)]^{2n+1}\quad,\quad D_{\Was}(E,p)=\sum_{n=0}^{\infty}\delta^{2n+1}[D(E,p)]^{2(n+1)},
\end{equation}
where
\begin{equation}
    D(E,p)=\frac{1}{\gamma-\sqrt{\frac{3}{4}p^2-M_NE-i\epsilon}}.
\end{equation}
In the Wigner-limit $\delta=0$ and it becomes readily apparent that only the Wigner-symmetric component of the dibaryon matrix, Eq.~\eqref{eq:dibmatrix}, remains.  Similar to dibaryon propagators the three-nucleon vertex function can also be expanded in powers of $\delta$ giving
\begin{equation}
    \Gc_{\Ws,m}(E_B,p)=\sum_{n=0}^\infty \Gc_{\Ws,m}^{(2n)}(E_B,p)\delta^{2n}\quad,\quad \Gc_{\Was,m}(E_B,p)=\sum_{n=0}^{\infty}\Gc_{\Was,m}^{(2n+1)}(E_B,p)\delta^{2n+1}.
\end{equation}
As shown in Ref.~\cite{Vanasse:2016umz} we can redefine the vertex function as
\begin{align}
    &\Gct_{\Ws,m}^{(2n)}(E_B,p)=\Gc_{\Ws,m}^{(2n)}(E_B,p)+D(E_B,p)\Gct_{\Was,m}^{(2n-1)}(E_B,p)\\\nonumber
    &\Gct_{\Was,m}^{(2n+1)}(E_B,p)=\Gc_{\Was,m}^{(2n+1)}(E_B,p)+D(E_B,p)\Gct_{\Ws,m}^{(2n)}(E_B,p),
\end{align}
which has the advantage of significantly simplifying the integral equations for the Wigner-SU(4) expanded vertex function.  Using this definition, expanding the integral equation for the LO vertex function in powers of $\delta$, and collecting like powers of $\delta$ gives the coupled set of integral equations~\cite{Vanasse:2016umz}
\begin{align}
    &\Gct_{\Ws,0}^{(2n)}(E_B,p)=2\delta_{0n}+D\left(E_B,p\right)\Gct_{\Was,0}^{(2n-1)}(E_B,p)+M(q,p,E_B)\otimes_q\Gct_{\Ws,0}^{(2n)}(E_B,q)\\\nonumber
    &\Gct_{\Was,0}^{(2n+1)}(E_B,p)=D\left(E_B,p\right)\Gct_{\Ws,0}^{(2n)}(E_B,p)-\frac{1}{2}M(q,p,E_B)\otimes_q\Gct_{\Was,0}^{(2n+1)}(E_B,q),
\end{align}
where
\begin{equation}
    M(q,p,E_B)=8\pi D\left(E_B,q\right)\frac{1}{qp}Q_0\left(\frac{q^2+p^2-M_NE_B}{qp}\right).
\end{equation}

Expanding the integral equation, Eq.~\eqref{eq:vertex}, for the NLO correction to the triton vertex function in powers of $\delta$ and collecting all $\mathcal{O}(\delta^0)$ contributions gives 
\begin{equation}
    \Gct_{\Ws,1}^{(0)}(E_B,p)=\frac{1}{2}\rho\left(\gamma+\sqrt{\frac{3}{4}p^2-M_NE_B}\,\right)\Gct_{\Ws,0}^{(0)}(E_B,p)+M(q,p,E_B)\otimes_q\Gct_{\Ws,1}^{(0)}(E_B,q).
\end{equation}
In this expression we also take the additional limit $\delta_r=0$.

\subsection{Inhomogeneous term for $E1$ amplitude}

\subsubsection{Leading-Order}

To rewrite the LO inhomogeneous term $\widetilde{\Bb}^{[0]}_{[0]}{}_{1\frac{1}{2}}^{\frac{1}{2}}(p,k)$ for the $E1$ moment in terms of the vertex function and dibaryon matrix in the Wigner-basis we make repeated use of the identity
\begin{equation}
    \left(\begin{array}{cc}
    1 & 0\\
    0 & 1
    \end{array}\right)=\frac{1}{2}\left(\begin{array}{rr}
    1 & 1\\
    -1 & 1
    \end{array}\right)\left(\begin{array}{rr}
    1 & -1\\
    1 & 1
    \end{array}\right).
\end{equation}
with Eq.~\eqref{eq:E1inhom} at LO to find
\begin{align}
     &\widetilde{\Bb}^{[0]}_{[0]}{}_{1\frac{1}{2}}^{\frac{1}{2}}(p,k)=-i\frac{e}{2M_N}\frac{2\tau_3}{\sqrt{3}\pi k_0}\int dqq^2\frac{1}{qp}\left[\left(\begin{array}{rr}
     1 & 0\\
     1 & 0
     \end{array}\right) pQ_0\left(\frac{q^2+p^2-M_NE-i\epsilon}{qp}\right)\right.\\\nonumber
    &\left.-\left(\begin{array}{rr}
    0 & 1\\
    0 & -1\end{array}\right) qQ_1\left(\frac{q^2+p^2-M_NE_B-i\epsilon}{qp}\right)\right]\mathbf{D}_W\left(E_B,q\right)\Gb_{W,0}(E_B,q)\\\nonumber
    &-i\frac{e}{2M_N}\frac{\tau_3}{2\sqrt{3} k_0}\Db^{-1}(E,p)\left(\begin{array}{cc} 1 & 0\\
    1 & 0
    \end{array}\right)\Db_W(E_B,p)p\Gb_{W,0}(E_B,p)\\\nonumber
    &+i\frac{e}{2M_N}\frac{\tau_3}{2\sqrt{3} k_0}p\left[2+\left(\begin{array}{rr}
    0 & 1\\
    0 & -1\end{array}\right) \Db^{-1}_W(E_B,p)\Db_W(E_B,p)\Gb_{W,0}(E_B,p)\right]\\\nonumber
\end{align}
Noting that
\begin{equation}
    \Db_W(E_B,p)\Gb_{W,0}(E_B,p)=D(E_B,p)\Gbt_{W,0}(E_B,p),
\end{equation}
where
\begin{equation}
    \Gbt_{W,n}(E_B,p)={\displaystyle\sum_{m=0}^{\infty}}\left(\begin{array}{l}
    \Gct_{\Ws,n}^{(2m)}(E_B,p)\\
    \Gct_{\Was,n}^{(2m+1)}(E_B,p)
    \end{array}\right),
\end{equation}
and
 \begin{equation}
     \Db^{-1}_{W}(E_B,p)=\left(D^{-1}(E_B,p)\left(\begin{array}{cc}
     1 & 0\\
     0 & 1\end{array}\right)+\delta\left(\begin{array}{cc}
     0 & 1\\
     1& 0\end{array}\right)\right),
 \end{equation}
 we expand the remaining vertex function in powers of $\delta$ and collect like powers of $\delta$ yielding

\begin{align}
     &\widetilde{\Bb}^{[0,n]}_{[0]}{}_{1\frac{1}{2}}^{\frac{1}{2}}(p,k)=-i\frac{e}{2M_N}\frac{2\tau_3}{\sqrt{3}\pi k_0}\int dqq^2\frac{1}{qp}\left[\left(\begin{array}{rr}
     1 & 0\\
     1 & 0
     \end{array}\right) pQ_0\left(\frac{q^2+p^2-M_NE-i\epsilon}{qp}\right)\right.\\\nonumber
    &\left.+\left(\begin{array}{rr}
    0 & 1\\
    0 & -1\end{array}\right) qQ_1\left(\frac{q^2+p^2-M_NE_B-i\epsilon}{qp}\right)\right]D\left(E_B,q\right)\Gbt^{(n)}_{W,0}(E_B,q)\\\nonumber
    &-i\frac{e}{2M_N}\frac{\tau_3}{2\sqrt{3} k_0}\Db^{-1}(E,p)\left(\begin{array}{cc} 1 & 0\\
    1 & 0
    \end{array}\right)D(E_B,p)p\Gbt^{(n)}_{W,0}(E_B,p)\\\nonumber
    &+i\frac{e}{2M_N}\frac{\tau_3}{2\sqrt{3} k_0}p\left[2 \delta_{n0}+\left(\begin{array}{rr}
    0 & 1\\
    0 & -1\end{array}\right) \Gbt^{(n)}_{W,0}(E_B,p)\right]\\\nonumber
    &+i\frac{e}{2M_N}\frac{\tau_3}{2\sqrt{3} k_0}p\left(\begin{array}{rr}
    1 & 0\\
    -1 & 0\end{array}\right) \delta D(E_B,p)\Gbt^{(n-1)}_{W,0}(E_B,p).\\\nonumber
\end{align}
Although $\Db^{-1}(E,p)$ contains powers of $\delta$ we do not expand them as they come from contributions after the photon.  The Wigner-SU(4) expanded inhomogeneous terms is notated as $\widetilde{\Bb}^{[0,n]}_{[0]}{}_{1\frac{1}{2}}^{\frac{1}{2}}(p,k)$ where the $[0,n]$ denotes LO in \EFT and $n$ denotes the order $\delta$ in the additional Wigner-SU(4) expansion.  Going to the Wigner-SU(4) limit it is apparent the inhomogeneous term for the $E1$ moment is not zero whereas it is known that the $M1$ moment is zero in this limit~\cite{Lin:2022yaf} in the zero-recoil limit.

\subsubsection{Next-to-Leading-Order}

Carrying out a similar expansion in powers of $\delta$ for the NLO inhomogeneous term while taking the limit $\delta_r=0$ gives the $\mathcal{O}(\delta^0)$ inhomogeneous term

\begin{align}
     &\widetilde{\Bb}^{[1,0]}_{[0]}{}_{1\frac{1}{2}}^{\frac{1}{2}}(p,k)=-i\frac{e}{2M_N}\frac{2\tau_3}{\sqrt{3}\pi k_0}\int dqq^2\frac{1}{qp}pQ_0\left(\frac{q^2+p^2-M_NE-i\epsilon}{qp}\right)\\\nonumber
    &\hspace{1cm}\times D\left(E_B,q\right)\Gct^{(0)}_{\Ws,1}(E_B,q)\left(\begin{array}{c}
    1\\
    1
    \end{array}\right)\\\nonumber
    &-i\frac{e}{2M_N}\frac{\tau_3}{2\sqrt{3} k_0}D(E_B,p)p\Gct^{(0)}_{\Ws,1}(E_B,p)\Db^{-1}(E,p)\left(\begin{array}{c} 1 \\
    1 
    \end{array}\right)\\\nonumber
    &+i\frac{e}{2M_N}\frac{\tau_3}{4\sqrt{3} k_0}\rho\left(\left(\begin{array}{cc}
    \gamma_t^2 & 0\\
    0 & \gamma_s^2 
    \end{array}\right)+\left(M_NE-\frac{3}{4}p^2\right)\right)D(E_B,p)p\Gct_{\Ws,0}^{(0)}(E_B,p)\left(\begin{array}{c}
    1 \\
    1 
    \end{array}\right).
\end{align}

\section{\label{sec:obs}Observables}

Including the $M1$ and $E1$ moment the relationship between the two-body triton photodisintegration amplitude in the spin and partial wave basis is given by
\begin{align}
    \label{eq:cross}
    &M_{m_1'm_2',\lambda m_2}(\vect{p},\vect{k})=\sum_{\alpha}\sqrt{4\pi}\CG{1}{\frac{1}{2}}{S'}{m_1'}{m_2'}{m_S'}\CG{L'}{S'}{J'}{m_{L'}}{m_S'}{M'}{Y_{L'}^{m_{L'}}}^{*}(\hat{\boldsymbol{p}})\\\nonumber
    &\left\{M_{[0]}{}^{J'}_{L'S'}(p,k)\epsilon_{\gamma}^m(\lambda)+M_{[1]}{}^{J'}_{L'S'}(p,k)\epsilon_{n\ell m}\hat{k}_n\epsilon_{\gamma}^\ell(\lambda)\right\}\CG{\frac{1}{2}}{1}{J'}{m_2}{m}{M'},
\end{align}
where $m_2$ is the triton spin, $\lambda$ the photon polarization, $m_1'$ ($m_2'$) the deuteron (nucleon) spin, and $\epsilon^{m}_{\gamma}(\lambda)$ is the $m$-th component of the photon polarization vector for polarization $\lambda$.  $\alpha$ sums over all quantum numbers and magnetic quantum numbers except $m_2$, $\lambda$, $m_1'$, and $m_2'$.  The two-body triton photodisintegration cross section is given by
\begin{align}
    &\sigma=\frac{1}{4}\frac{1}{2k_0}\sum_{m_1',m_2'}\sum_{m_2,\lambda}\int\frac{d^3p_n}{(2\pi)^3}\int\frac{d^3p_d}{(2\pi)^3}|M_{m_1'm_2',\lambda m_2}(p,k)|^2(2\pi)^4\boldsymbol{\delta}^3(\vect{p}_n+\vect{p}_d)\\\nonumber
    &\hspace{5cm}\times\delta\left(E_B+k_0-\frac{p_n^2}{2M_N}-\frac{p_d^2}{4M_N}+\frac{\gamma^2}{M_N}\right)
\end{align}
Plugging in Eq.~\eqref{eq:cross}, carrying out the integrals, and summing over nuclear spins and photon polarizations the two-body triton photodisintegration cross section is given by
\begin{equation}
    \sigma(\gamma t\to nd)=\frac{M_Np}{12\pi k_0}\sum_{L',S',J'}(2J'+1)\left\{|M_{[0]}{}^{J'}_{L'S'}(p,k)|^2+\frac{2}{3}|M_{[1]}{}^{J'}_{L'S'}(p,k)|^2\right\}.
\end{equation}
Expanding the amplitude perturbatively to NNLO yields
\begin{align}
    &\sigma(\gamma t\to nd)=\frac{M_Np}{12\pi k_0}\sum_{L',S',J'}(2J'+1)\\\nonumber
    &\times\left(\vphantom{\frac{2}{3}}\left\{|M_{[0]}^{[0]}{}^{J'}_{L'S'}(p,k)+M_{[0]}^{[1]}{}^{J'}_{L',S'}(p,k)|^2+2\mathrm{Re}\left[M_{[0]}^{[0]}{}^{J'}_{L'S'}(p,k)\left(M_{[0]}^{[2]}{}^{J'}_{L'S'}(p,k)\right)^*\right]\right\}\right.\\\nonumber
&\left.+\frac{2}{3}\left\{|M_{[1]}^{[0]}{}^{J'}_{L'S'}(p,k)+M_{[1]}^{[1]}{}^{J'}_{L',S'}(p,k)|^2+2\mathrm{Re}\left[M_{[1]}^{[0]}{}^{J'}_{L'S'}(p,k)\left(M_{[1]}^{[2]}{}^{J'}_{L'S'}(p,k)\right)^*\right]\right\}\right).
\end{align}

Another observable of interest is the photon polarization asymmetry $R_c$ in cold $nd$ capture.  In the asymmetry the neutron is polarized and the difference in cross sections for different outgoing photon polarizations is taken.  The asymmetry is defined by
\begin{align}
R_c\hat{\boldsymbol{z}}\cdot\widehat{\boldsymbol{k}}=\frac{\displaystyle\sum_{m_1'}\sum_{m_2,\lambda}\lambda|M_{m_1'\frac{1}{2},\lambda m_2}(\vect{p},\vect{k})|^2}{\displaystyle\sum_{m_1'}\sum_{m_2,\lambda}|M_{m_1'm_2',\lambda m_2}(\vect{p},\vect{k})|^2}.
\end{align}
Plugging in Eq.~\eqref{eq:cross} and explicitly summing over quantum numbers, the asymmetry $R_c$ is given by
\begin{equation}
    \label{eq:Rc}
    R_c=\frac{-4\mathrm{Re}\left[\left(M_{[1]}{}^{\frac{3}{2}}_{0\frac{3}{2}}\right)^*M_{[1]}{}^{\frac{1}{2}}_{0\frac{1}{2}}\right]-\left|M_{[1]}{}^{\frac{1}{2}}_{0\frac{1}{2}}\right|^2+5\left|M_{[1]}{}^{\frac{3}{2}}_{0\frac{3}{2}}\right|^2}{3\left|M_{[1]}{}^{\frac{1}{2}}_{0\frac{1}{2}}\right|^2+6\left|M_{[1]}{}^{\frac{3}{2}}_{0\frac{3}{2}}\right|^2}.
\end{equation}
The amplitudes in the expressions for $R_C$ in principle contain contributions from all orders in \EFT.  Thus these amplitudes in the expression for $R_C$ should be expanded perturbatively to NNLO and the lengthy expression is given in App.~\ref{sec:appRC}.  However, for asymmetries with terms in their denominator a relatively small change to terms in the denominator can lead to relatively large changes in the asymmetry.  Thus to accurately predict such asymmetries one must go to high orders in a strictly perturbative expansion.  To overcome this we treat the numerator perturbatively while resumming amplitudes in the denominator to NNLO giving what we term the partially resummed expression that is also given in App.~\ref{sec:appRC}.   We also look at the expression where we resum the amplitudes to \NNLO in both the numerator and denominator of $R_c$ given in App.~\ref{sec:appRC}.

\section{\label{sec:res}Results}

\subsection{Physical Limit ($\delta_r=0$, $\delta\neq 0$)}

\begin{figure}[hbt]
    \centering
    \includegraphics[width=150mm]{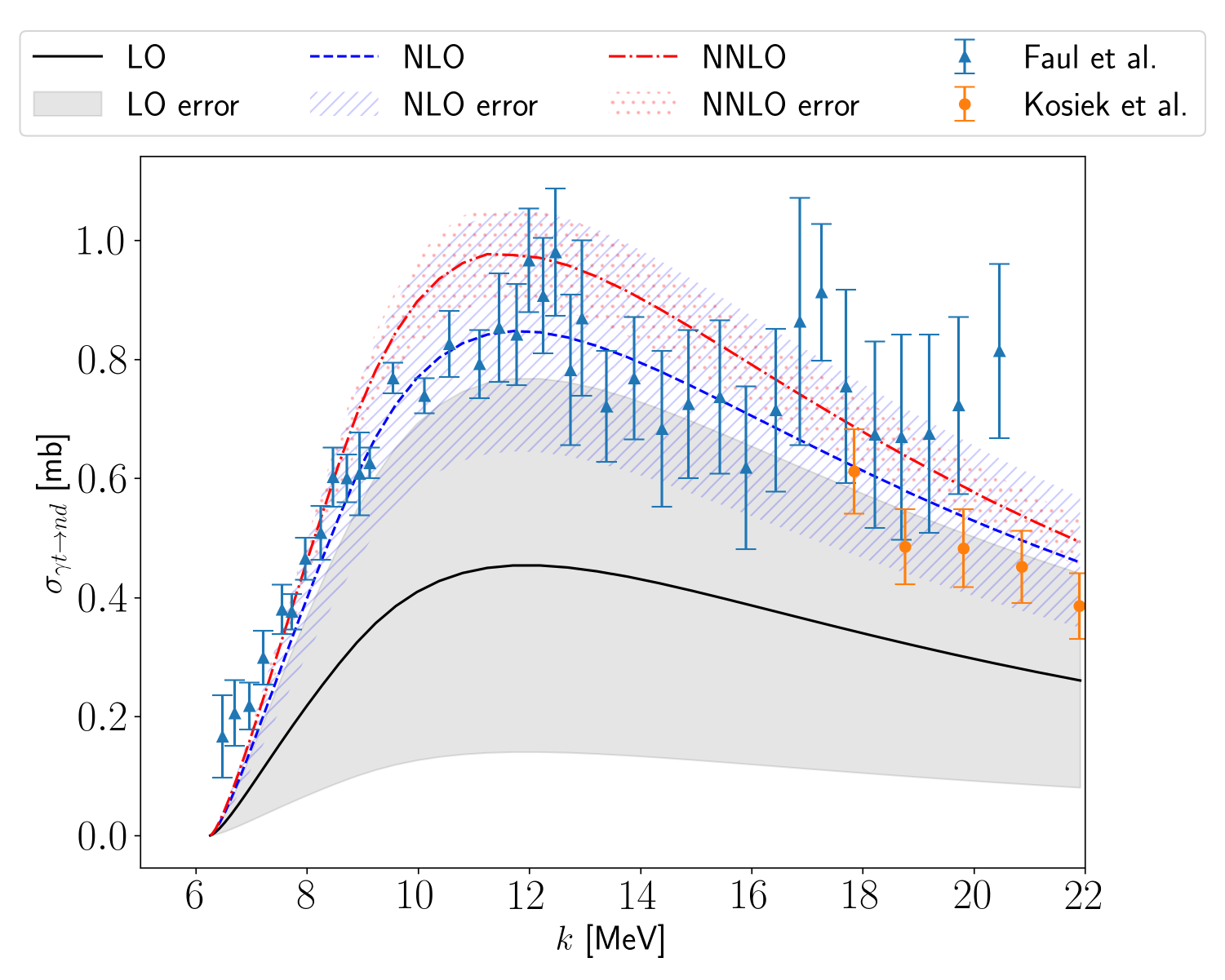}
    \caption{Two-body photodisintegration cross section of the triton as a function of photon momentum.  The LO \EFT prediction is shown with a solid line and its associated naive error band with a  solid grey band.  NLO and NNLO \EFT predictions are shown with a blue dashed line and red dashed dotted line respectively.  The error bands at NLO are shown with blue hatched lines and at NNLO with red dots.  The experimental data from Faul et al.~\cite{Faul:1981zz} is shown with triangle points and the experimental data at higher photon energies from Kosiek et al.~\cite{KOSIEK1966199} is shown with circles.}
    \label{fig:cross}
\end{figure}

The two-body triton photodisintegration cross section as a function of photon momentum at LO, NLO, and NNLO is shown in Fig.~\ref{fig:cross}.
Error bands about each order are given by the naive error estimate from \EFT, $Q\sim(Z_t-1)/2=0.3454$.  However, the cross section is given by the amplitude squared and therefore this error estimate is multiplied by a factor of two.  The error bands are $\approx$ 69\%, 24\%, and 8\% at LO, NLO, and NNLO respectively. Experimental data comes from Faul et al.~\cite{Faul:1981zz} and at higher photon energies from Kosiek et al.~\cite{KOSIEK1966199}.  In general there seems to be good overlap between the experimental values and \EFT predictions.  Near threshold the NNLO \EFT results underpredict the Faul data while between roughly 10 to 16~MeV NNLO \EFT consistently overpredicts the Faul data except near the peak of the Faul data.  Above $\approx$16~MeV the Faul data trends above the NNLO \EFT results while the Kosiek data trends below.  However, this is approaching the \EFT breakdown scale, $k_0=3\Lambda_{\cancel{\pi}}^2/(4M_N)+E_B-\gamma_t^2/M_N\approx22$~MeV, where the \EFT results have more inherent uncertainty.  Note, the value of the $M1$ moment depends on how $L_1^{(m)}$ and $L_2^{(m)}$ are fit.  For details of how to fit these LECs consult Ref.~\cite{Lin:2022yaf}.  Fitting $L_2^{(m)}$ to the deuteron magnetic moment, $L_1^{(0)}$ to the triton magnetic moment ($\mu_{\jjvH}$), and $L_1^{(1)}$ to the cold $np$ capture cross section ($\sigma_{np}$) gives the results in Fig.~\ref{fig:cross}.  Fitting to different data as in Ref.~\cite{Lin:2022yaf} for $L_1^{(m)}$ has little discernible effect for the plot in Fig.~\ref{fig:cross} since the $E1$ moment quickly dominates over the $M1$ moment at the photon energies shown.

At low energies the asymmetry $R_C$ is dominated by the $M1$ moment.  Fitting $L_1^{(0)}$ to $\mu_{\jjvH}$ and $L_1^{(1)}$ to $\sigma_{np}$~\cite{Lin:2022yaf} gives the \EFT predictions for $R_C$ up to \NNLO in 
Table~\ref{tab:fit1}.
\begin{table}
\begin{tabular}{|c|cccc|}
\hline
& $\sigma_{nd}$ [mb] & $R_C$[pert] & $R_C$[resum] & $R_C$ [presum] \\\hline
LO & 0.314(217) & -0.252(272) & -0.252(272) & -0.252(272) \\
NLO & 0.393[164] & -0.662(217)[471] &-0.484(20)[74] & -0.518(54)[87]\\
NNLO & 0.447[130] & -0.359(81)[352] & -0.446(14)[122] & -0.417(41)[140]\\\hline
Expt. &0. 508$\pm$ 0.015\cite{Jurney:1982zz} & -0.42$\pm$ 0.03\cite{Konijnenberg:1988gq} & & \\\hline
\end{tabular}
\caption{Values of $R_C$ and $\sigma_{nd}$ compared to experiment. $L_1^{(0)}$ is fit to $\mu_{\jjvH}$ and $L_1^{(1)}$ is fit to $\sigma_{np}$. The strictly perturbative (pert), fully resummed (resum), and partially resummed (presum) results, in which only the amplitudes in the denominator are resummed, are shown.\label{tab:fit1}}
\end{table}
For completeness we also show the values of $\sigma_{nd}$ calculated previously in Ref.~\cite{Lin:2022yaf}.  The errors shown in parentheses are using naive error estimates from \EFT while the errors in square brackets come from propagating the uncertainty of the $L_1^{(0)}$ and $L_1^{(1)}$ LECs.  Further details of how these uncertainty estimates are obtained can be found in App.~\ref{sec:apperr} and Ref.~\cite{Lin:2022yaf}.  Table~\ref{tab:fit2} contains the same information as Table~\ref{tab:fit1} except $L_1^{(0)}$ is simultaneously fit to $\sigma_{nd}$, $\sigma_{np}$, and $\mu_{\jjvH}$ while $L_1^{(1)}$ is simultaneously fit to $\sigma_{nd}$ and $\sigma_{np}$. In each table $R_C$ is calculated using a full perturbative (pert) expansion, by resumming (resum) the amplitudes to NNLO in both numerator and denominator, and by only  resumming the amplitudes to NNLO in the denominator, while perturbatively expanding the numerator known as partially resumming (presum).

For $R_C$ we find nearly all our results agree with the experimental value within theoretical and experimental errors.  However, the NLO predictions for the presum and resum results for both fits of $L_1^{(0)}$ and $L_1^{(1)}$ considered disagree with experiment within naive theoretical errors.  We note that the change from LO to NLO and NLO to \NNLO for $R_C$ seems larger than what one would naively expect from the \EFT expansion.  This is in part because the $M1$ amplitudes in the Wigner-SU(4) limit are zero.  Expanding in both \EFT and the Wigner-SU(4) breaking parameter $\delta$, it was argued in Ref~\cite{Lin:2022yaf} that LO $\mathcal{O}(\delta^2)$ and \NNLO $\mathcal{O}(\delta^0)$ terms would both be leading terms in this dual expansion.  As a result a naive \EFT expansion does not work well for the $M1$ amplitude.  Another factor leading to large changes from order to order is that $R_C$ is very sensitive to the values of its denominator in Eq.~\eqref{eq:Rc}. It is readily apparent that the perturbative results vary much more from order to order than either the partially resummed or fully resummed results.  This is because the denominator for $R_C$ is perturbatively expanded for the perturbative results  whereas for the partially resummed and fully resummed results the denominator is fully resummed.  Thus order by order the denominator for the partially resummed and fully resummed results get closer to the physical value of the denominator, while for the perturbative results the denominator only includes the LO amplitudes and corrections beyond LO to the denominator are included via a perturbative expansion.  The fact that a small change in the denominator of $R_C$ can lead to large changes in $R_C$ and that the $M1$ amplitudes change more than the naive \EFT expansion from order to order conspire to make a much more sizable change for the perturbative results form order to order.  As a result we take the fully resummed result as the prediction of this work for $R_C$.

\begin{table}
\begin{tabular}{|c|cccc|}
\hline
& $\sigma_{nd}$ [mb] & $R_C$[pert] & $R_C$[resum] & $R_C$ [presum] \\\hline
LO & 0.314(217) & -0.252(272) & -0.252(272) & -0.252(272) \\
NLO & 0.480(114) & -0.891(514) &-0.500(1) & -0.532(47)\\
NNLO & 0.511(42) & -0.0304(5743) & -0.441(15) & -0.403(58) \\\hline
Expt. &0. 508$\pm$ 0.015\cite{Jurney:1982zz} & -0.42$\pm$ 0.03\cite{Konijnenberg:1988gq} & & \\\hline
\end{tabular}
\caption{\label{tab:fit2} Identical to Table~\ref{tab:fit1} except $L_1^{(0)}$ ($L_1^{(1)}$) fit to $\sigma_{np}$, $\sigma_{nd}$, and $\mu_{\jjvH}$ at NLO ($\sigma_{np}$ and $\sigma_{nd}$ at \NNLO).}
\end{table}

\subsection{\label{subsec:QSU(4)}Quasi Wigner-SU(4) expansion}

\begin{figure}[hbt]
    \centering
    \includegraphics[width=150mm]{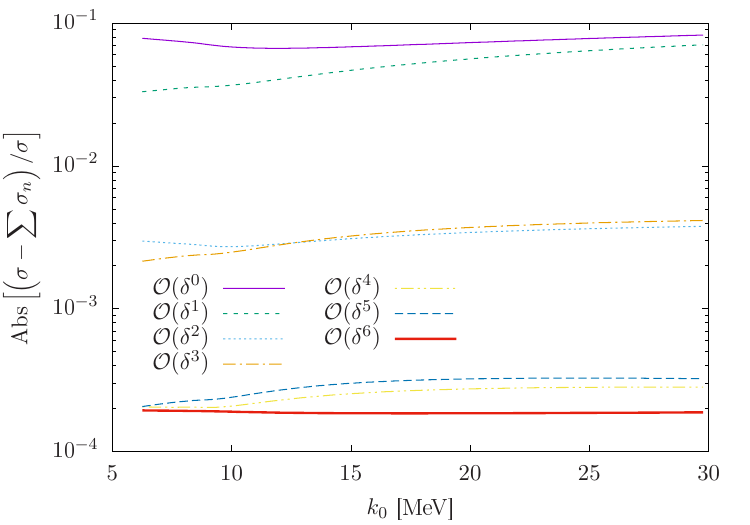}
    \caption{Absolute value of the difference between the two-body triton photo-disintegration cross section at order $\delta^n$ in the quasi Wigner-SU(4) expansion ($\sigma_n$) and the two-body triton photo-disintegration cross section at LO in \EFT with no quasi Wigner-SU(4) expansion ($\sigma$) normalised by $\sigma$.  This is shown as a function of photon energy all the way up to $\mathcal{O}(\delta^6)$.}
    \label{fig:conv}
\end{figure}

\begin{figure}[hbt]
    \centering
    \includegraphics[width=150mm]{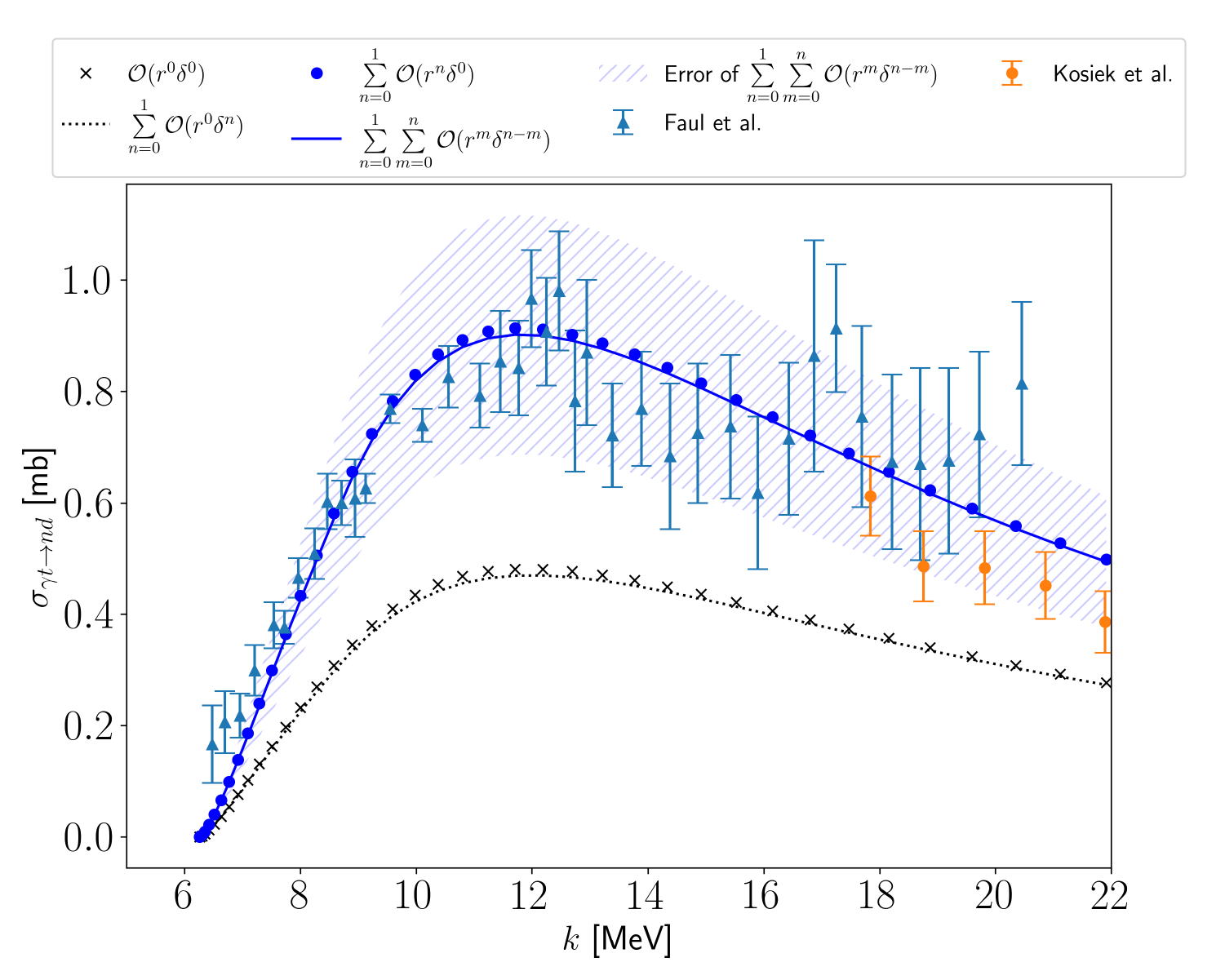}
    \caption{Two-body photodisintegration cross section of the triton as a function of photon momentum in the quasi Wigner-SU(4) expansion.  The $\mathcal{O}(r^0\delta^0)$ \EFT prediction is shown with a dotted line, the $\mathcal{O}(r^0\delta^0)+\mathcal{O}(r^0\delta^1)$ \EFT prediction is shown with $\times$'s, the $\mathcal{O}(r^0\delta^0)+\mathcal{O}(r^1\delta^0)$ \EFT prediction with large blue dots, and finally the $\mathcal{O}(r^0\delta^0)+\mathcal{O}(r^0\delta^1)+\mathcal{O}(r^1\delta^0)$ \EFT prediction with a solid blue line.  For clarity error bands are only shown for the $\mathcal{O}(r^0\delta^0)+\mathcal{O}(r^0\delta^1)+\mathcal{O}(r^1\delta^0)$ result.  The experimental data from Faul et al.~\cite{Faul:1981zz} is shown with triangle points and the experimental data at higher photon energies from Kosiek et al.~\cite{KOSIEK1966199} is shown with circles.}
    \label{fig:Wignerexp}
\end{figure}

To evaluate the effectiveness and utility of the quasi Wigner-SU(4) expansion in Fig.~\ref{fig:conv} we plot the difference between the LO two-body triton photodisintegration cross section in the quasi Wigner-SU(4) expansion at different orders of $\delta^n$, denoted $\sigma_n$, and the LO triton photodisintegration cross section with no Wigner-SU(4) expansion, denoted $\sigma$. This difference is normalized by $\sigma$.  Note, these cross sections only contain contributions from $E1$ as contributions from $M1$ are negligible on the plot and carrying out the quasi Wigner-SU(4) expansion for $M1$ is more complicated.  Order by order convergence in the quasi Wigner-SU(4) expansion can be clearly seen in Fig~\ref{fig:conv} thus demonstrating the effectiveness of the quasi Wigner-SU(4) expansion. Figure~\ref{fig:Wignerexp} shows the $\mathcal{O}(r^0\delta^0)$, $\mathcal{O}(r^0\delta^0)+\mathcal{O}(r^0\delta^1)$, $\mathcal{O}(r^0\delta^0)+\mathcal{O}(r^1\delta^0)$, and $\mathcal{O}(r^0\delta^0)+\mathcal{O}(r^0\delta^1)+\mathcal{O}(r^1\delta^0)$ contribution to the two-body triton photodisintegration cross section in the quasi Wigner-SU(4) expansion, where $r$ refers to range corrections and its order gives the \EFT expansion order.  The $\mathcal{O}(\delta)$ correction is observed to be a small correction and the $\mathcal{O}(r^0\delta^0)+\mathcal{O}(r^0\delta^1)+\mathcal{O}(r^1\delta^0)$ term agrees well with experimental data just like our NLO results not in the quasi Wigner-SU(4) expansion.

The quasi Wigner-SU(4) expansion for the $M1$ moment is complicated due to the three-body force in the outgoing \DS channel.  In the quasi Wigner-SU(4) expansion, to use the same three-body force for both the vertex function and scattering amplitudes, Wigner-SU(4) corrections would have to be added for the triton binding energy at each order.  This is a rather involved process.  Instead we take a simpler less rigorous approach as in Ref.~\cite{Lin:2022yaf} which promotes the energy dependent three-body force and the three-nucleon magnetic moment counterterm Eq.~\eqref{eq:magNNLO} to LO, in a dual \EFT Wigner-SU(4) expansion, giving the modified LO inhomogeneous term for the $M1$ moment
\begin{align}
    \label{eq:inhomwigLO}
    &\widetilde{\Bb}^{[0,0]}_{W[1]}{}_{1\frac{1}{2}}^{\frac{1}{2}}(p,k)=\frac{\delta}{\sqrt{3}k_0}4\tau_3\kappa_1\left(\!\!\begin{array}{rr}
    0 & 1\\
    -1 & 0
    \end{array}\!\!\right)\Db\left(E_B,p\right)\Gb_{0}(E_B,p)+\sqrt{3}H_{\mathrm{LO}}\Sigma_0'(E_B)(\mu_t-\mu_p)\oneb.
\end{align}
$\mu_t$ ($\mu_p$) is the triton (proton) magnetic moment.  Although this has all the correct contributions at LO in the dual \EFT Wigner-SU(4) expansion, it also has higher-order corrections in powers of $\delta$ contained in the vertex function and dibaryon matrix.   Using this inhomogeneous term we find the results for $R_C$ and $\sigma_{nd}$ shown in Table~\ref{tab:fit3}.  With this expansion we see that the LO predictions are very close to the experimental predictions for $\sigma_{nd}$ and $R_C$.

\begin{table}
\begin{tabular}{|c|cc|}
\hline
& $\sigma_{nd}$ [mb] & $R_C$[pert]  \\\hline
LO(mod) & 0.511(217) & -0.422(272)  \\
Expt. &0. 508$\pm$ 0.015\cite{Jurney:1982zz} & -0.42$\pm$ 0.03\cite{Konijnenberg:1988gq} \\\hline
\end{tabular}
\caption{\label{tab:fit3} Values of $R_C$ and $\sigma_{nd}$ compared to experiment using Eq.~\eqref{eq:inhomwigLO} for the inhomogeneous term in the integral equation.}
\end{table}

\section{\label{sec:conc}Conclusion}
In this work we calculated the $E1$ moment in addition to the $M1$ moment for two-body triton photodisintegration up to NNLO in \EFT.  Using this we calculated the two-body triton photodisintegration cross section as a function of photon momentum to NNLO in \EFT finding good agreement with experiment.  We also calculated the polarization asymmetry $R_C$ in cold $nd$ capture.  Fitting $L_1^{(0)}$ to the triton magnetic moment and $L_1^{(1)}$ to cold $np$  capture we found $R_C=-0.446(14)[122]$ at NNLO in \EFT, while for fitting $L_1^{(0)}$ simultaneously to cold $np$ capture, cold $nd$ capture, and the triton magnetic moment and $L_1^{(1)}$ simultaneously to cold $np$ capture and cold $nd$ capture we found $R_C=-0.441(15)$ at NNLO in \EFT.  These predictions agree within both naive (values in parentheses) and propagated errors (values in brackets) with the experimental value of $R_C=-0.42\pm0.03$~\cite{Konijnenberg:1988gq}.  For details of of how errors were determined consult App.~\ref{sec:apperr}.

We also considered the consequences of Wigner-SU(4) symmetry for these observables.  For the $E1$ moment of two-body triton photon disintegration the outgoing states are $P$-wave and thus contain no three-body force.  Thus we were able to perform a Wigner-SU(4) expansion of the incoming triton state while not performing the Wigner-SU(4) expansion for the outgoing $nd$ scattering state after the photon.  This expansion we termed the quasi Wigner-SU(4) expansion.  Carrying out this expansion we find that the two-body triton photodisintegration cross section as a function of photon momentum again agrees with experiment.  We also demonstrate that the two-body triton photon disintegration cross section in the quasi Wigner-SU(4) expansion converges towards the cross section in the physical limit.  Carrying out a similar analysis for the $M1$ moment is complicated by the three-body force in the outgoing \DS $nd$ scattering state after the photon.  To treat the three-body force consistently for both bound and scattering states while carrying out the quasi Wigner-SU(4) expansion, Wigner-SU(4) corrections must be added to the triton binding energy complicating the analysis of the $M1$ moment in the quasi Wigner-SU(4) expansion.  Instead, we carry out a less rigorous quasi Wigner-SU(4) expansion for the $M1$ moment by promoting the energy dependent three-body force and the three-nucleon magnetic moment counterterm to LO in a dual \EFT Wigner-SU(4) counting scheme.  This promotion gives all the correct contributions at LO in the dual expansion but contains an infinite subset of higher order terms.  Doing this we calculate the $M1$ moment and found a photon polarization asymmetry of $R_C=-0.422(272)$ again in good agreement with experiment.

In all our calculations we chose the effective ranges in the \TS and \SSp channels to be equal ($\delta_r=0$).  Corrections to this are approximately N$^3$LO in \EFT and thus this approximation does not affect the uncertainty of our NNLO calculations.  As found in Ref.~\cite{Lin:2022yaf}, choosing $\delta_r\neq0$ at NNLO in \EFT, there appears to be either a slow convergence or divergence in the cutoff dependence of the $M1$ moment and this is also observed for the $E1$ moment in this work.  To figure out if it is a slow convergence or divergence we could in principle calculate $M1$ and $E1$ to larger cutoffs but are limited by numerical issues.  A detailed asymptotic analysis could also answer if this is a slow convergence or divergence.  Future work should address this issue.

Future work should consider the electric quadrupole moment ($E2$) which is necessary to accurately determine certain polarization asymmetries in $nd$ capture.  The $E2$ moment may also explain why the experimental data is slightly larger than our prediction for the two-body triton photodisintegration cross section near the two-body breakup threshold.  Near this threshold the $M1$ and $E2$ moments become more important.  Building upon this work we also want to consider parity-violation in $nd$ capture which is a possible future experiment~\cite{Alarcon:2023gfu}.

\acknowledgments{We thank Roxanne Springer for useful discussions.  XL is supported by the Henry W. Newson fellowship for Fall 2023, by the U.S. Department of Energy, Office of Science, Office of Nuclear Physics, under Award Number DE-FG02-05ER41368 and DE-SC0024622, and by the NSF grant PHY-2044632.}

\appendix

\section{Simplifying the $E1$ moment\label{app:simpler}}

By using the integral equation for the vertex function, Eq.~\eqref{eq:vertex}, and shifting the integral equation, Eq.~\eqref{eq:Tamp}, for the $E1$ moment the expression for the integral equation of the $E1$ moment can be considerably simplified.  Similarly this can be done for the $M1$ moment, for details of the simplification of $M1$ consult Ref.~\cite{Lin:2022yaf}.  Summing diagrams (a) through (f) at the appropriate order (See Figs.~\ref{fig:inhomLO}, \ref{fig:inhomNLO}, and \ref{fig:inhomNNLO}) the inhomogeneous term for the $E_1$ moment up to LO, NLO, and NNLO is given by
\begin{align}
    &{\Bb}^{[n]}_{[0]}{}_{1\frac{1}{2}}^{\frac{1}{2}}(p,k)=-i\frac{e}{2M_N}\frac{1}{\sqrt{3}\pi k_0}\left(\begin{array}{cc}
    -(1-\tau_3) & (3+\tau_3)\\
    (3+\tau_3) & -\frac{1}{3}(3+5\tau_3)
    \end{array}\right)\int dqq^2\mathbf{D}\left(E_B,q\right)\Gb_n(E_B,q)\\\nonumber
    &\frac{1}{qp}\left[qQ_1\left(\frac{q^2+p^2-M_NE-i\epsilon}{qp}\right)+pQ_0\left(\frac{q^2+p^2-M_NE-i\epsilon}{qp}\right)\right.\\\nonumber
    &\left.-qQ_1\left(\frac{q^2+p^2-M_NE_B-i\epsilon}{qp}\right)-pQ_0\left(\frac{q^2+p^2-M_NE_B-i\epsilon}{qp}\right)\right]\\\nonumber
    &+i\frac{M_N}{2\sqrt{3}}\frac{e}{2M_N}\left(\begin{array}{cc}
    4 & 0\\
    0 & (4+\frac{8}{3}\tau_3)
    \end{array}\right)\mathbf{D}\left(E_B,p\right)\Gb_n(E_B,p)\\\nonumber
    &\frac{1}{2}p\left(\frac{1}{\sqrt{\frac{3}{4}p^2-M_NE_B-i\epsilon}+\sqrt{\frac{3}{4}p^2-M_NE-i\epsilon}}\right)\\\nonumber
    &-i\frac{1}{\pi\sqrt{3}}\frac{e}{2M_N}\left(\begin{array}{cc}
    -(1+\tau_3) & (3-\tau_3)\\
    3(1+\tau_3) & -\frac{1}{3}(3-\tau_3)
    \end{array}\right)\int dqq^2\mathbf{D}\left(E_B,q\right)\Gb_n(E_B,q)\frac{q}{k_0}\\\nonumber
    &\frac{1}{qp}Q_1\left(\frac{q^2+p^2-M_NE_B-i\epsilon}{qp}\right)\\\nonumber
    &+i\left(\begin{array}{cc}
    1+\tau_3 & 0\\
    0 &\frac{1}{3}(3-\tau_3)
    \end{array}\right)\frac{e}{2M_N}\left[\vphantom{\left(\begin{array}{rr}
    1 & -3\\
    -3 & 1
    \end{array}\right)}\delta_{n0}\oneb-\frac{1}{\pi}\int_0^{\Lambda}dqq^2\frac{1}{qp}Q_0\left(\frac{q^2+p^2-M_NE-i\epsilon}{qp}\right)\right.\\\nonumber
    &\left.\left(\begin{array}{rr}
    1 & -3\\
    -3 & 1
    \end{array}\right)\Db\left(E_B,q\right)\Gb_n(E_B,q)\right]\frac{p}{\sqrt{3}k_0}\\\nonumber
    &-i\frac{1}{\sqrt{3}}\frac{e}{2M_N}p\sum_{m=0}^{n-1}\left(\begin{array}{cc}
    c_{0t}^{(m)} & 0\\
    0 & \frac{1}{3}c_{0s}^{(m)}(3+2\tau_3)
    \end{array}\right)\mathbf{D}\left(E_B,p\right)\Gb_{n-1-m}(E_B,p).
\end{align}
We first simplify this expression by collecting like terms.  Then using Eq.~\eqref{eq:vertex} to simplify some of the integrals with the vertex function yields
\begin{align}
     &{\Bb}^{[n]}_{[0]}{}_{1\frac{1}{2}}^{\frac{1}{2}}(p,k)=-i\frac{e}{2M_N}\frac{2\tau_3}{\sqrt{3}\pi k_0}\int dqq^2\frac{1}{qp}\left[\Mcb pQ_0\left(\frac{q^2+p^2-M_NE-i\epsilon}{qp}\right)\right.\\\nonumber
    &\left.-\Mcb^T qQ_1\left(\frac{q^2+p^2-M_NE_B-i\epsilon}{qp}\right)\right]\mathbf{D}\left(E_B,q\right)\Gb_n(E_B,q)\\\nonumber
    &-i\frac{e}{2M_N}\frac{1}{\sqrt{3}\pi k_0}\left(\begin{array}{cc}
    -(1-\tau_3) & (3+\tau_3)\\
    (3+\tau_3) & -\frac{1}{3}(3+5\tau_3)
    \end{array}\right)\int dqq^2\mathbf{D}\left(E_B,q\right)\Gb_n(E_B,q)\\\nonumber
    &\frac{1}{qp}qQ_1\left(\frac{q^2+p^2-M_NE-i\epsilon}{qp}\right)\\\nonumber
    &+i\frac{e}{2M_N}\frac{1}{2\sqrt{3} k_0}\left(\begin{array}{cc}
    2+\tau_3 & \tau_3\\
    -\tau_3 & 2+\frac{1}{3}\tau_3
    \end{array}\right)p\Gb_n(E_B,p)\\\nonumber
    &-i\frac{e}{2M_N}\frac{1}{2\sqrt{3} k_0}\sum_{m=1}^{n}\left(\begin{array}{cc}
    2+\tau_3 & \tau_3\\
    -\tau_3 & 2+\frac{1}{3}\tau_3
    \end{array}\right)p\mathbf{R}_m\left(E_B,p\right)\Gb_{n-m}(E_B,p)\\\nonumber
    &+i\frac{M_N}{2\sqrt{3}}\frac{e}{2M_N}\left(\begin{array}{cc}
    4 & 0\\
    0 & (4+\frac{8}{3}\tau_3)
    \end{array}\right)\mathbf{D}\left(E_B,p\right)\Gb_n(E_B,p)\\\nonumber
    &\frac{1}{2}p\left(\frac{1}{\sqrt{\frac{3}{4}p^2-M_NE_B-i\epsilon}+\sqrt{\frac{3}{4}p^2-M_NE-i\epsilon}}\right)\\\nonumber
    &+i\frac{e}{2M_N}\frac{\tau_3}{2\sqrt{3} k_0}\Mcb p\delta_{n0}\oneb\\\nonumber
    &-i\frac{1}{\sqrt{3}}\frac{e}{2M_N}p\sum_{m=0}^{n-1}\left(\begin{array}{cc}
    c_{0t}^{(m)} & 0\\
    0 & \frac{1}{3}c_{0s}^{(m)}(3+2\tau_3)
    \end{array}\right)\mathbf{D}\left(E_B,p\right)\Gb_{n-1-m}(E_B,p).
\end{align}
We also collected all terms with $\oneb$.  Next we shift the definition of the integral equation, Eq.~\eqref{eq:Tamp}, to absorb part of the inhomogeneous term into the amplitude $\Tb^{[n]}_{[0]}{}_{L'S'}^{J'}(p,k)$. Shifting the integral equation gives
\begin{align}
    \label{eq:Ttildeamp}
&\widetilde{\Tb}^{[n]}_{[0]}{}_{L'S'}^{J'}(p,k)=\Tb^{[n]}_{[0]}{}_{L'S'}^{J'}(p,k)\\\nonumber
&\hspace{2cm}-i\frac{e}{2M_N}\frac{1}{2\sqrt{3} k_0}\Db^{-1}(E,p)\left(\begin{array}{cc}
    2+\tau_3 & -\tau_3\\
    \tau_3 & 2+\frac{1}{3}\tau_3
    \end{array}\right)\Db(E_B,p)p\Gb_n(E_B,p).   
\end{align}
When the outgoing legs are on shell for $\widetilde{\Tb}^{[n]}_{[0]}{}_{L'S'}^{J'}(p,k)$ the amplitudes ${\Tb}^{[n]}_{[0]}{}_{L'S'}^{J'}(p,k)$ and $\widetilde{\Tb}^{[n]}_{[0]}{}_{L'S'}^{J'}(p,k)$ only differ in the outgoing nucleon spin-singlet dibaryon channel.  However, for two-body triton photodisintegration we only look at the outgoing nucleon spin-triplet dibaryon channel and therefore $\widetilde{\Tb}^{[n]}_{[0]}{}_{L'S'}^{J'}(p,k)$ and ${\Tb}^{[n]}_{[0]}{}_{L'S'}^{J'}(p,k)$ are equivalent. Note, for three-body triton photdisintegration these are no longer equivalent. Using Eq.~\eqref{eq:Tamp} and Eq.~\eqref{eq:Ttildeamp} the new inhomogeneous term of the integral equation for $\widetilde{\Tb}^{[n]}_{[0]}{}_{L'S'}^{J'}(p,k)$ is
\begin{align}
     &\widetilde{\Bb}^{[n]}_{[0]}{}_{1\frac{1}{2}}^{\frac{1}{2}}(p,k)=-i\frac{e}{2M_N}\frac{2\tau_3}{\sqrt{3}\pi k_0}\int dqq^2\frac{1}{qp}\left[\Mcb pQ_0\left(\frac{q^2+p^2-M_NE-i\epsilon}{qp}\right)\right.\\\nonumber
    &\left.-\Mcb^T qQ_1\left(\frac{q^2+p^2-M_NE_B-i\epsilon}{qp}\right)\right]\mathbf{D}\left(E_B,q\right)\Gb_n(E_B,q)\\\nonumber
    &-i\frac{e}{2M_N}\frac{1}{2\sqrt{3} k_0}\Db^{-1}(E,p)\left(\begin{array}{cc}
    2+\tau_3 & -\tau_3\\
    \tau_3 & 2+\frac{1}{3}\tau_3
    \end{array}\right)\Db(E_B,p)p\Gb_n(E_B,p)\\\nonumber
    &+i\frac{e}{2M_N}\frac{1}{2\sqrt{3} k_0}\sum_{m=1}^{n}\mathbf{R}_{m}\left(E,p\right)\Db^{-1}(E,p)\left(\begin{array}{cc}
    2+\tau_3 & -\tau_3\\
    \tau_3 & 2+\frac{1}{3}\tau_3
    \end{array}\right)\Db(E_B,p)p\Gb_{n-m}(E_B,p)\\\nonumber
    &+i\frac{e}{2M_N}\frac{1}{2\sqrt{3} k_0}\left(\begin{array}{cc}
    2+\tau_3 & \tau_3\\
    -\tau_3 & 2+\frac{1}{3}\tau_3
    \end{array}\right)p\Gb_n(E_B,p)\\\nonumber
    &-i\frac{e}{2M_N}\frac{1}{2\sqrt{3} k_0}\sum_{m=1}^{n}\left(\begin{array}{cc}
    2+\tau_3 & \tau_3\\
    -\tau_3 & 2+\frac{1}{3}\tau_3
    \end{array}\right)p\mathbf{R}_m\left(E_B,p\right)\Gb_{n-m}(E_B,p)\\\nonumber
    &+i\frac{M_N}{2\sqrt{3}}\frac{e}{2M_N}\left(\begin{array}{cc}
    4 & 0\\
    0 & (4+\frac{8}{3}\tau_3)
    \end{array}\right)\mathbf{D}\left(E_B,p\right)\Gb_n(E_B,p)\\\nonumber
    &\frac{1}{2}p\left(\frac{1}{\sqrt{\frac{3}{4}p^2-M_NE_B-i\epsilon}+\sqrt{\frac{3}{4}p^2-M_NE-i\epsilon}}\right)\\\nonumber
    &+i\frac{e}{2M_N}\frac{\tau_3}{2\sqrt{3} k_0}\Mcb p\delta_{n0}\oneb\\\nonumber
    &-i\frac{1}{\sqrt{3}}\frac{e}{2M_N}p\sum_{m=0}^{n-1}\left(\begin{array}{cc}
    c_{0t}^{(m)} & 0\\
    0 & \frac{1}{3}c_{0s}^{(m)}(3+2\tau_3)
    \end{array}\right)\mathbf{D}\left(E_B,p\right)\Gb_{n-1-m}(E_B,p).
\end{align}

Using the identity
\begin{equation}
    \frac{M_Nk_0}{\sqrt{\frac{3}{4}p^2-M_NE_B-i\epsilon}+\sqrt{\frac{3}{4}p^2-M_NE-i\epsilon}}=\Db^{-1}(E,p)-\Db^{-1}(E_B,p),
\end{equation}
and 

\begin{align}
    \left(\begin{array}{cc}
    c_{0t}^{(m-1)} & 0\\
    0 & c_{0s}^{(m-1)}\end{array}\right)=\frac{1}{k_0}\left(\mathbf{R}_m(E,p)\Db^{-1}(E,p)-\mathbf{R}_m(E_B,p)\Db^{-1}(E_B,p)\right),
\end{align}
the time ordered contributions before and after the photon interactions for diagram-(e) and diagram-(f) respectively, in Figs.~\ref{fig:inhomLO} through \ref{fig:inhomNNLO} are made explicit.  With these identities the inhomogeneous term can be simplified to Eq.~\eqref{eq:E1inhom} using the definition of $\Mcb$ in Eq.~\eqref{eq:Mcb}.

\section{\label{sec:apperr}Error Analysis}
\subsection{General formalism}
Consider an observable $\mathcal{O}(A)$ as a smooth function of the amplitude $A$. \footnote{The error analysis here is motivated and based on the fact that, while the exact form of $\mathcal{O}(A)$ as an explicit function of $A$ can be often easily obtained,  the LECs as well as the form of $A$ as a function of the LECs are only known perturbatively and/or numerically.} The EFT expansion of $A$ at the $m$-th order is given by the series $A_m = \sum_{n = 0}^{m}a_n$, where $a_n$ is the $n$-th order correction to $A$ and depends on LECs up to the $n$-th EFT order. The naive EFT uncertainty of $A_m$ is defined as
\begin{align}
    \Delta_{\text{N}} \left( A_m\right) =  \lvert Q^{m+1}A_m\rvert,
\end{align}
where the subscript ``N" on $\Delta_{\text{N}}$ indicates that this is the naive EFT error estimate.
We also define the propagated uncertainty of $A_m$ as that caused by the uncertainty of a LEC $C_m$, denoted as $\Delta C_m$, that first appears at $m$-th EFT order:
\begin{align}
    \Delta_{\text{P}} \left( A_m\right) =  \left| \frac{\partial A_m}{\partial C_m}\right|\Delta C_m,
\end{align}
where the subscript ``P" on $\Delta_{\text{P}}$ indicates that this is the error propagated through the LEC $C_m$. The value of $C_m$ is fit to another observable $\mathcal{O'}$ with an experimental value $\mathcal{O'}^{\text{exp}}$, and  $\Delta C_m$ is determined from the $m$-th order naive EFT error of $\mathcal{O'}$
\begin{align}
    \Delta C_m =  \left|\frac{\beta Q^{m+1}\mathcal{O'}^{\text{exp}}}{\partial \mathcal{O'}/\partial C_m } \right|
\end{align}
where $\beta = 1$  ($\beta = 2$) if $\mathcal{O'}$ is proportional to an amplitude (amplitude squared).

The uncertainty of $\mathcal{O}(A_m)$ depends on how $\mathcal{O}$ is expanded in $A_m$. If we resum $\mathcal{O}(A_m)$ as a function of $A_m$ and use the notation\footnote{This differs from, for example, the resummation of the effective ranges in the dibaryon propagators. One reason is that the resummation of the effective ranges may affect the renormalization of $A_m$ at a given EFT order, while expanding (or not expanding) $\mathcal{O}$ as a function of $A_m$ obviously has no impact on such renormalization.}
\begin{align}
    &\mathcal{O}^{\text{resum}}_{m} = \mathcal{O}(A_m),
\end{align}
the naive [propagated] uncertainty of $\mathcal{O}^{\text{resum}}_{m}$ is obtained by expanding the difference between $\mathcal{O}^{\text{resum}}_{m+1}$ and $\mathcal{O}^{\text{resum}}_{m}$ to the lowest order in $a_{m+1}$ and estimating $\left|a_{m+1}\right|$ using $\Delta_{\text{N}}(A_{m+1})$ [$\Delta_{\text{P}}(A_m)$]: 
\begin{align}
    \Delta_{\text{N/P}}\left(\mathcal{O}^{\text{resum}}_{m} \right) \approx \Delta_{\text{N/P}}(A_m)\left|\mathcal{O}^{(1)}(A_m)\right|, 
    \label{eq:error-Omresum}
\end{align}
where $\mathcal{O}^{(k)}$ is the $k$-th derivative of $\mathcal{O}$ with respect to $A$

On the other hand, the uncertainty is different for $\mathcal{O}(A_m)$ expanded strictly perturbatively. For example, at $m = 1$ and $2$ we have
\begin{align}
    &\mathcal{O}(A_1) = \mathcal{O}(a_0+ a_1) \approx \mathcal{O}(a_0) + a_1\mathcal{O}^{(1)}(a_0) \equiv \mathcal{O}^{\text{pert}}_{1},
    \label{eq:O1pert}
\end{align}
and 
\begin{align}
    &\mathcal{O}(A_2) = \mathcal{O}(a_0+ a_1+ a_2) \approx\mathcal{O}(a_0) +\left(a_1+a_2\right)\mathcal{O}^{(1)}(a_0) + a_1^2\frac{\mathcal{O}^{(2)}(a_0)}{2}\equiv \mathcal{O}^{\text{pert}}_{2},
    \label{eq:O2pert}
\end{align}
respectively. To extract the naive (propagated) uncertainty of $\mathcal{O}^{\text{pert}}_{1}$, we compare $\mathcal{O}^{\text{pert}}_{1}$ with $\mathcal{O}^{\text{pert}}_{2}$ and estimate $\left|a_2\right|$ using $\Delta_{\text{N}}(A_1)$ ($\Delta_{\text{P}}(A_1)$), which gives
\begin{align}
    \Delta_{\text{N/P}}\left(\mathcal{O}^{\text{pert}}_{1} \right) = \sqrt{\left(\Delta_{\text{N/P}}(A_1)\mathcal{O}^{(1)}(a_0)\right)^2 + \left(a_1^2\frac{\mathcal{O}^{(2)}(a_0)}{2}\right)^2},
    \label{eq:error-O1pert}
\end{align}
where $a_2\mathcal{O}^{(1)}(a_0)$ and $a_1^2\frac{\mathcal{O}^{(2)}(a_0)}{2}$ are treated as independent uncertainties because their relative phase is not informed by the naive estimate of $\left|a_2\right|$. Specifically, the first term under the square root in Eq.~\eqref{eq:error-O1pert} originates from the EFT-expansion truncation error (due to the lack of knowledge of NNLO and higher-order LECs) as well as the perturbative expansion of $A$ in NLO LECs (due to the lack of knowledge of the exact dependence of $A$ on NLO LECs), while the second term originates from the perturbative expansion of $\mathcal{O}$ around $a_0$ (there is no lack of knowledge of how $\mathcal{O}$ depends on $A$).  Similarly, expanding $\mathcal{O}(A_3)$ strictly perturbatively gives
\begin{align}
    \mathcal{O}(A_3) & = \mathcal{O}(a_0 + a_1 + a_2+a_3)\\ \nonumber
    & \approx  \mathcal{O}(a_0) +\left(a_1+a_2+a_3\right)\mathcal{O}^{(1)}(a_0) + \left(a_1^2 + 2a_1a_2\right)\frac{\mathcal{O}^{(2)}(a_0)}{2} + a_1^3\frac{\mathcal{O}^{(3)}(a_0)}{6}\\ \nonumber &
    \equiv \mathcal{O}^{\text{pert}}_{3}.
\end{align}
Comparing $\mathcal{O}^{\text{pert}}_{2}$ with $\mathcal{O}^{\text{pert}}_{3}$ and estimating $\left|a_3\right|$ using $\Delta_{\text{N}}(A_2)$ ($\Delta_{\text{P}}(A_2)$) gives the naive (propagated) uncertainty of $\mathcal{O}^{\text{pert}}_{2}$
\begin{align}
    \Delta_{\text{N/P}}\left(\mathcal{O}^{\text{pert}}_{2} \right) = \sqrt{\left(\Delta_{\text{N/P}}(A_2)\mathcal{O}^{(1)}(a_0)\right)^2 + \left(a_1a_2\mathcal{O}^{(2)}(a_0) + a_1^3\frac{\mathcal{O}^{(3)}(a_0)}{6}\right)^2},
    \label{eq:error-O2pert}
\end{align}
which can also be generalized beyond \NNLO using the same procedure.

The uncertainty of the partially resummed $\mathcal{O}(A_m)$ depends on the choice of partial re-summation and is obtained in a similar manner to that of $\mathcal{O}^{\text{pert}}_{m}$. For example, assuming  $\mathcal{O}(A_m)$ can be written as a ratio
\begin{align}
     \mathcal{O}(A_m) = \frac{F(A_m)}{G(A_m)},
\end{align}
with a partial re-summation defined by
\begin{align}
    \mathcal{O}^{\text{presum}}_m \equiv \frac{F^{\text{pert}}_m}{G^{\text{resum}}_m},
\end{align}
where $F^{\text{pert}}_m$ and $G^{\text{pert}}_m$ are defined similarly as $\mathcal{O}^{\text{pert}}_m$ and $\mathcal{O}^{\text{resum}}_m$, respectively.  The difference between $\mathcal{O}^{\text{presum}}_2$ and $\mathcal{O}^{\text{presum}}_1$ is given by
\begin{align}
   \mathcal{O}^{\text{presum}}_2 - \mathcal{O}^{\text{presum}}_1   & \equiv \frac{F^{\text{pert}}_2}{G^{\text{resum}}_2} - \frac{F^{\text{pert}}_1}{G^{\text{resum}}_1} \\\nonumber
   & =  \frac{\left(F^{\text{pert}}_2 - F^{\text{pert}}_1\right)G^{\text{resum}}_1 - F^{\text{pert}}_1\left(G^{\text{resum}}_2 - G^{\text{resum}}_1\right)}{G^{\text{resum}}_2G^{\text{resum}}_1}\\\nonumber
   & \approx  a_2\mathcal{O}^{\text{presum}}_1 \left(\frac{{F}^{(1)}(a_0)}{F^{\text{pert}}_1} - \frac{{G}^{(1)}(A_1)}{G^{\text{resum}}_1}\right)
 + \frac{a_1^2}{2}\frac{{F}^{(2)}(a_0)}{G^{\text{resum}}_1}.
\end{align}
To obtain the last line above,  we used Eqs.~\eqref{eq:O1pert} and~\eqref{eq:O2pert} on $F^{\text{pert}}_{1/2}$, approximated
\begin{align}
    G^{\text{resum}}_2 = G(A_2) \approx G(A_1) + a_2 G^{(0)}(A_1),
\end{align}
and kept terms linear in $a_2$ or quadratic in $a_1$.
The uncertainty of $ \mathcal{O}^{\text{presum}}_1$ is thus given by
\begin{align}
     \Delta_{\text{N/P}}\left(\mathcal{O}^{\text{presum}}_{1} \right) = \sqrt{\left(\Delta_{\text{N/P}}(A_1)\mathcal{O}^{\text{presum}}_1 \left(\frac{\mathcal{F}^{(1)}(a_0)}{F^{\text{pert}}_1} - \frac{\mathcal{G}^{(1)}(A_1)}{G^{\text{resum}}_1}\right)\right)^2 + \left( \frac{a_1^2}{2}\frac{\mathcal{F}^{(2)}(a_0)}{G^{\text{resum}}_1}\right)^2}.
\end{align}

\subsection{Naive and propagated uncertainty for $R_c$ }
The error analysis above can be generalized to observables that depend on more than one amplitude-like variable. For instance, $R_c$ depends on both the $nd$ capture amplitude in the \DS channel and that in the \QS channel, denoted $A$ and $B$ below for simplicity with \EFT expansions $A_m = \sum_{n = 0}^{m}a_n$ and $B_m = \sum_{n = 0}^{m}b_n$. To obtain the naive EFT uncertainty of $R_c$, the naive EFT uncertainties for the amplitudes in these two channels, are considered independent and added quadratically.  The naive EFT uncertainty of $R_c^{\text{LO}}$ is given by\footnote{Here $A$ and $B$ are taken to be purely real, which is a very good approximation for neutrons in cold $nd$ capture.}
\begin{align}
    \Delta_{\text{N}} \left(R_c^{\text{LO}}\right) &=\sqrt{\left[\left(\frac{\partial R_c}{\partial A}\right)(a_0,b_0)\right]^2(Qa_0)^2 + \left[\left(\frac{\partial R_c}{\partial B}\right)(a_0,b_0)\right]^2(Qb_0)^2}  \\\nonumber
    &=\sqrt{(-0.192)^2 + 0.762^2} = 0.272
\end{align}

In contrast, when calculating the propagated uncertainty of $R_c$, $A$ and $B$ are not necessarily independent variables as both of them could depend on some LECs used to propagate the uncertainty at a given EFT order. In our calculation of $R_c$ at NLO [NNLO], only $L_1^{(0)}$ [$L_1^{(1)}$] is used for this purpose. For example, the propagated error for the re-summed $R_c^{\text{NNLO}}$ is given by
\begin{align}
    \Delta_{\text{P}} \left(R_c^{\text{NNLO}}\right) &=\Delta_{\text{P}}\left(L_1^{(1)}\right)\left|\left(\frac{\partial R_c}{\partial A}\frac{\partial A}{\partial L_1^{(1)}}+ \frac{\partial R_c}{\partial B}\frac{\partial B}{\partial L_1^{(1)}}\right)(A_2, B_2) \right|\\\nonumber
    &=1.01[\text{fm}]\times 0.121[\text{fm}^{-1}]\\\nonumber
    & = 0.122
\end{align}

\section{\label{sec:appRC}Expansion of $R_C$}

Expanding the amplitudes perturbatively for the polarization asymmetry $R_c$ to \NNLO yields
\begin{align}
    &R_c=\left\{-4\mathrm{Re}\left[\left(M_{[1]}^{[0]}{}^{\frac{3}{2}}_{0\frac{3}{2}}\right)^*M_{[1]}^{[0]}{}^{\frac{1}{2}}_{0\frac{1}{2}}\right]-\left|M_{[1]}^{[0]}{}^{\frac{1}{2}}_{0\frac{1}{2}}\right|^2+5\left|M_{[1]}^{[0]}{}^{\frac{3}{2}}_{0\frac{3}{2}}\right|^2-4\mathrm{Re}\left[\left(M_{[1]}^{[0]}{}^{\frac{3}{2}}_{0\frac{3}{2}}\right)^*M_{[1]}^{[1]}{}^{\frac{1}{2}}_{0\frac{1}{2}}\right]\right.\\\nonumber
    &-4\mathrm{Re}\left[\left(M_{[1]}^{[1]}{}^{\frac{3}{2}}_{0\frac{3}{2}}\right)^*M_{[1]}^{[0]}{}^{\frac{1}{2}}_{0\frac{1}{2}}\right]-2\mathrm{Re}\left[M_{[1]}^{[0]}{}^{\frac{1}{2}}_{0\frac{1}{2}}\left(M_{[1]}^{[1]}{}^{\frac{1}{2}}_{0\frac{1}{2}}\right)^*\right]+10\mathrm{Re}\left[M_{[1]}^{[0]}{}^{\frac{3}{2}}_{0\frac{3}{2}}\left(M_{[1]}^{[1]}{}^{\frac{3}{2}}_{0\frac{3}{2}}\right)^*\right]\\\nonumber
    &-4\mathrm{Re}\left[\left(M_{[1]}^{[0]}{}^{\frac{3}{2}}_{0\frac{3}{2}}\right)^*M_{[1]}^{[2]}{}^{\frac{1}{2}}_{0\frac{1}{2}}\right]-4\mathrm{Re}\left[\left(M_{[1]}^{[2]}{}^{\frac{3}{2}}_{0\frac{3}{2}}\right)^*M_{[1]}^{[0]}{}^{\frac{1}{2}}_{0\frac{1}{2}}\right]-4\mathrm{Re}\left[\left(M_{[1]}^{[1]}{}^{\frac{3}{2}}_{0\frac{3}{2}}\right)^*M_{[1]}^{[1]}{}^{\frac{1}{2}}_{0\frac{1}{2}}\right]\\\nonumber
    &\left.-2\mathrm{Re}\left[M_{[1]}^{[0]}{}^{\frac{1}{2}}_{0\frac{1}{2}}\left(M_{[1]}^{[2]}{}^{\frac{1}{2}}_{0\frac{1}{2}}\right)^*\right]+10\mathrm{Re}\left[M_{[1]}^{[0]}{}^{\frac{3}{2}}_{0\frac{3}{2}}\left(M_{[1]}^{[2]}{}^{\frac{3}{2}}_{0\frac{3}{2}}\right)^*\right]-\left|M_{[1]}^{[1]}{}^{\frac{1}{2}}_{0\frac{1}{2}}\right|^2+5\left|M_{[1]}^{[1]}{}^{\frac{3}{2}}_{0\frac{3}{2}}\right|^2\right\}/\\\nonumber
    &\left\{3\left|M_{[1]}^{[0]}{}^{\frac{1}{2}}_{0\frac{1}{2}}\right|^2+6\left|M_{[1]}^{[0]}{}^{\frac{3}{2}}_{0\frac{3}{2}}\right|^2\right\}\\\nonumber
    &+\left\{-4\mathrm{Re}\left[\left(M_{[1]}^{[0]}{}^{\frac{3}{2}}_{0\frac{3}{2}}\right)^*M_{[1]}^{[0]}{}^{\frac{1}{2}}_{0\frac{1}{2}}\right]-\left|M_{[1]}^{[0]}{}^{\frac{1}{2}}_{0\frac{1}{2}}\right|^2+5\left|M_{[1]}^{[0]}{}^{\frac{3}{2}}_{0\frac{3}{2}}\right|^2-4\mathrm{Re}\left[\left(M_{[1]}^{[0]}{}^{\frac{3}{2}}_{0\frac{3}{2}}\right)^*M_{[1]}^{[1]}{}^{\frac{1}{2}}_{0\frac{1}{2}}\right]\right.\\\nonumber
    &\left.-4\mathrm{Re}\left[\left(M_{[1]}^{[1]}{}^{\frac{3}{2}}_{0\frac{3}{2}}\right)^*M_{[1]}^{[0]}{}^{\frac{1}{2}}_{0\frac{1}{2}}\right]-2\mathrm{Re}\left[M_{[1]}^{[0]}{}^{\frac{1}{2}}_{0\frac{1}{2}}\left(M_{[1]}^{[1]}{}^{\frac{1}{2}}_{0\frac{1}{2}}\right)^*\right]+10\mathrm{Re}\left[M_{[1]}^{[0]}{}^{\frac{3}{2}}_{0\frac{3}{2}}\left(M_{[1]}^{[1]}{}^{\frac{3}{2}}_{0\frac{3}{2}}\right)^*\right]\right\}\times\\\nonumber
    &\left\{-6\mathrm{Re}\left[\left(M_{[1]}^{[1]}{}^{\frac{1}{2}}_{0\frac{1}{2}}\right)^*M_{[1]}^{[0]}{}^{\frac{1}{2}}_{0\frac{1}{2}}\right]-12\mathrm{Re}\left[\left(M_{[1]}^{[1]}{}^{\frac{3}{2}}_{0\frac{3}{2}}\right)^*M_{[1]}^{[0]}{}^{\frac{3}{2}}_{0\frac{3}{2}}\right]\right\}/\\\nonumber
    &\left\{3\left|M_{[1]}^{[0]}{}^{\frac{1}{2}}_{0\frac{1}{2}}\right|^2+6\left|M_{[1]}^{[0]}{}^{\frac{3}{2}}_{0\frac{3}{2}}\right|^2\right\}^2\\\nonumber
    &+\left\{-4\mathrm{Re}\left[\left(M_{[1]}^{[0]}{}^{\frac{3}{2}}_{0\frac{3}{2}}\right)^*M_{[1]}^{[0]}{}^{\frac{1}{2}}_{0\frac{1}{2}}\right]-\left|M_{[1]}^{[0]}{}^{\frac{1}{2}}_{0\frac{1}{2}}\right|^2+5\left|M_{[1]}^{[0]}{}^{\frac{3}{2}}_{0\frac{3}{2}}\right|^2\right\}\times\\\nonumber
    &\left\{-6\mathrm{Re}\left[\left(M_{[1]}^{[2]}{}^{\frac{1}{2}}_{0\frac{1}{2}}\right)^*M_{[1]}^{[0]}{}^{\frac{1}{2}}_{0\frac{1}{2}}\right]-12\mathrm{Re}\left[\left(M_{[1]}^{[2]}{}^{\frac{3}{2}}_{0\frac{3}{2}}\right)^*M_{[1]}^{[0]}{}^{\frac{3}{2}}_{0\frac{3}{2}}\right]-3\left|M_{[1]}^{[1]}{}^{\frac{1}{2}}_{0\frac{1}{2}}\right|^2-6\left|M_{[1]}^{[1]}{}^{\frac{3}{2}}_{0\frac{3}{2}}\right|^2\right\}/\\\nonumber
    &\left\{3\left|M_{[1]}^{[0]}{}^{\frac{1}{2}}_{0\frac{1}{2}}\right|^2+6\left|M_{[1]}^{[0]}{}^{\frac{3}{2}}_{0\frac{3}{2}}\right|^2\right\}^2\\\nonumber
    &+\left\{-4\mathrm{Re}\left[\left(M_{[1]}^{[0]}{}^{\frac{3}{2}}_{0\frac{3}{2}}\right)^*M_{[1]}^{[0]}{}^{\frac{1}{2}}_{0\frac{1}{2}}\right]-\left|M_{[1]}^{[0]}{}^{\frac{1}{2}}_{0\frac{1}{2}}\right|^2+5\left|M_{[1]}^{[0]}{}^{\frac{3}{2}}_{0\frac{3}{2}}\right|^2\right\}\times\\\nonumber
    &\left\{-6\mathrm{Re}\left[\left(M_{[1]}^{[1]}{}^{\frac{1}{2}}_{0\frac{1}{2}}\right)^*M_{[1]}^{[0]}{}^{\frac{1}{2}}_{0\frac{1}{2}}\right]-12\mathrm{Re}\left[\left(M_{[1]}^{[1]}{}^{\frac{3}{2}}_{0\frac{3}{2}}\right)^*M_{[1]}^{[0]}{}^{\frac{3}{2}}_{0\frac{3}{2}}\right]\right\}^2/\\\nonumber
    &\left\{3\left|M_{[1]}^{[0]}{}^{\frac{1}{2}}_{0\frac{1}{2}}\right|^2+6\left|M_{[1]}^{[0]}{}^{\frac{3}{2}}_{0\frac{3}{2}}\right|^2\right\}^3
\end{align}
Perturbatively expanding the numerator of $R_C$ while resumming the denominator to NNLO yields
\begin{align}
    &R_c=\left\{-4\mathrm{Re}\left[\left(M_{[1]}^{[0]}{}^{\frac{3}{2}}_{0\frac{3}{2}}\right)^*M_{[1]}^{[0]}{}^{\frac{1}{2}}_{0\frac{1}{2}}\right]-\left|M_{[1]}^{[0]}{}^{\frac{1}{2}}_{0\frac{1}{2}}\right|^2+5\left|M_{[1]}^{[0]}{}^{\frac{3}{2}}_{0\frac{3}{2}}\right|^2-4\mathrm{Re}\left[\left(M_{[1]}^{[0]}{}^{\frac{3}{2}}_{0\frac{3}{2}}\right)^*M_{[1]}^{[1]}{}^{\frac{1}{2}}_{0\frac{1}{2}}\right]\right.\\\nonumber
    &-4\mathrm{Re}\left[\left(M_{[1]}^{[1]}{}^{\frac{3}{2}}_{0\frac{3}{2}}\right)^*M_{[1]}^{[0]}{}^{\frac{1}{2}}_{0\frac{1}{2}}\right]-2\mathrm{Re}\left[M_{[1]}^{[0]}{}^{\frac{1}{2}}_{0\frac{1}{2}}\left(M_{[1]}^{[1]}{}^{\frac{1}{2}}_{0\frac{1}{2}}\right)^*\right]+10\mathrm{Re}\left[M_{[1]}^{[0]}{}^{\frac{3}{2}}_{0\frac{3}{2}}\left(M_{[1]}^{[1]}{}^{\frac{3}{2}}_{0\frac{3}{2}}\right)^*\right]\\\nonumber
    &-4\mathrm{Re}\left[\left(M_{[1]}^{[0]}{}^{\frac{3}{2}}_{0\frac{3}{2}}\right)^*M_{[1]}^{[2]}{}^{\frac{1}{2}}_{0\frac{1}{2}}\right]-4\mathrm{Re}\left[\left(M_{[1]}^{[2]}{}^{\frac{3}{2}}_{0\frac{3}{2}}\right)^*M_{[1]}^{[0]}{}^{\frac{1}{2}}_{0\frac{1}{2}}\right]-4\mathrm{Re}\left[\left(M_{[1]}^{[1]}{}^{\frac{3}{2}}_{0\frac{3}{2}}\right)^*M_{[1]}^{[1]}{}^{\frac{1}{2}}_{0\frac{1}{2}}\right]\\\nonumber
    &\left.-2\mathrm{Re}\left[M_{[1]}^{[0]}{}^{\frac{1}{2}}_{0\frac{1}{2}}\left(M_{[1]}^{[2]}{}^{\frac{1}{2}}_{0\frac{1}{2}}\right)^*\right]+10\mathrm{Re}\left[M_{[1]}^{[0]}{}^{\frac{3}{2}}_{0\frac{3}{2}}\left(M_{[1]}^{[2]}{}^{\frac{3}{2}}_{0\frac{3}{2}}\right)^*\right]-\left|M_{[1]}^{[1]}{}^{\frac{1}{2}}_{0\frac{1}{2}}\right|^2+5\left|M_{[1]}^{[1]}{}^{\frac{3}{2}}_{0\frac{3}{2}}\right|^2\right\}/\\\nonumber
    &\left\{3\left|\sum_{n=0}^{2}M_{[1]}^{[n]}{}^{\frac{1}{2}}_{0\frac{1}{2}}\right|^2+6\left|\sum_{n=0}^2M_{[1]}^{[n]}{}^{\frac{3}{2}}_{0\frac{3}{2}}\right|^2\right\}.
\end{align}
For completeness we also look at the expression where we resum the amplitudes to \NNLO in both the numerator and denominator of $R_c$ giving the fully resummed expression
\begin{equation}
    R_c=\frac{-4\mathrm{Re}\left[\displaystyle\sum_{n=0}^2\left(M_{[1]}^{[n]}{}^{\frac{3}{2}}_{0\frac{3}{2}}\right)^*\displaystyle\sum_{m=0}^2M_{[1]}^{[m]}{}^{\frac{1}{2}}_{0\frac{1}{2}}\right]-\left|\displaystyle\sum_{n=0}^2M_{[1]}^{[n]}{}^{\frac{1}{2}}_{0\frac{1}{2}}\right|^2+5\left|\displaystyle\sum_{n=0}^2M_{[1]}^{[n]}{}^{\frac{3}{2}}_{0\frac{3}{2}}\right|^2}{3\left|\displaystyle\sum_{n=0}^2M_{[1]}^{[n]}{}^{\frac{1}{2}}_{0\frac{1}{2}}\right|^2+6\left|\displaystyle\sum_{n=0}^2M_{[1]}^{[n]}{}^{\frac{3}{2}}_{0\frac{3}{2}}\right|^2}.
\end{equation}

\bibliography{ref}

\end{document}